\documentclass{aa}
\usepackage{natbib,psfig,graphicx}
\def\mathnew{\mathsurround=0pt}
\def\simov#1#2{\lower .5pt\vbox{\baselineskip0pt \lineskip-.5pt
\ialign{$\mathnew#1\hfil##\hfil$\crcr#2\crcr\sim\crcr}}}

\def\MeV{Me\kern-0.11em V}
\def\keV{ke\kern-0.11em V}

\begin{document}

\title{Very deep spectroscopy of the Coma cluster line of sight:
exploring new territories 
\thanks{Based on observations collected at the European Organisation for
Astronomical Research in the Southern Hemisphere, Chile (program:
081.A-0172).  Also based on observations obtained with
MegaPrime/MegaCam, a joint project of CFHT and CEA/DAPNIA, at the
Canada-France-Hawaii Telescope (CFHT) which is operated by the National
Research Council (NRC) of Canada, the Institut National des Sciences de
l'Univers of the Centre National de la Recherche Scientifique (CNRS) of
France, and the University of Hawaii. This work is also partly based on
data products produced at TERAPIX and the Canadian Astronomy Data Centre
as part of the Canada-France-Hawaii Telescope Legacy Survey, a
collaborative project of NRC and CNRS. }}

\author{C.~Adami\inst{1} \and 
V.~Le Brun\inst{1} \and 
A.~Biviano\inst{2} \and 
F.~Durret\inst{3} \and 
F.~Lamareille\inst{4} \and 
R.~Pell\'o\inst{4} \and 
O.~Ilbert\inst{1} \and 
A.~Mazure\inst{1} \and 
R.~Trilling\inst{3} \and
M.P.~Ulmer\inst{5} 
}

\offprints{C. Adami \email{christophe.adami@oamp.fr}}

\institute{
LAM, OAMP, Universit\'e Aix-Marseille $\&$ CNRS, P\^ole de l'Etoile, Site de
Ch\^ateau Gombert, 38 rue Fr\'ed\'eric Joliot-Curie,
13388 Marseille 13 Cedex, France
\and
INAF-Osservatorio Astronomico di Trieste, via G. B. Tiepolo 11, I-34143 Trieste, Italy
\and
Institut d'Astrophysique de Paris, CNRS, UMR~7095, Universit\'e Pierre et 
Marie Curie, 98bis Bd Arago, 75014 Paris, France 
\and
Laboratoire d'Astrophysique de Toulouse-Tarbes, Universit\'e de Toulouse,
CNRS, 14 Av. Edouard Belin, 
31400 Toulouse, France
\and
Department of Physics $\&$ Astronomy, Northwestern University, 2131
Sheridan Road,
Evanston, IL 60208-2900, USA
}

\date{Accepted . Received ; Draft printed: \today}

\authorrunning{Adami et al.}

\titlerunning{Very deep spectroscopy of the Coma cluster line of sight:
exploring new territories}

\abstract 
{Environmental effects are known to have an important influence on
cluster galaxies, but studies at very faint magnitudes (R$>21$) are
almost exclusively based on imaging. We present here a very deep
spectroscopic survey of galaxies on the line of sight to the Coma
cluster.}
{After a series of papers based on deep multi--band imaging of the Coma
cluster, we explore spectroscopically part of the central regions of
Coma, in order to confirm and generalize previous results, concerning in
particular the galaxy luminosity function, red sequence, stellar
populations and the most likely formation scenario for the Coma
cluster.}
{We have obtained reliable VIMOS redshifts for 715 galaxies in the direction
of the Coma cluster centre in the unprecedented magnitude range
$21\leq$R$\leq 23$, corresponding to the absolute magnitude range
$-14\leq {\rm M_R} \leq -12$.}
{We confirm the substructures previously identified in Coma by Adami et
al. (2005a), and identify three new substructures. We detect a large
number of groups behind Coma, in particular a large structure at
z$\sim$0.5, the SDSS Great Wall, and a large and very young previously
unknown structure at z$\sim$0.054, which we named the background massive
group (BMG). These structures account for the mass maps derived from a
recent weak lensing analysis by Gavazzi et al. (2009). The orbits of
dwarf galaxies are probably anisotropic and radial, and could originate
from field galaxies radially falling into the cluster along the numerous
cosmological filaments surrounding Coma. Spectral characteristics of
Coma dwarf galaxies show that red or absorption line galaxies have
larger stellar masses and are older than blue or emission line
galaxies. R$\leq$22 galaxies show less prominent absorption lines than
R$\geq$22 galaxies. This trend is less clear for field galaxies which
are similar to R$\geq$22 Coma galaxies. This suggests that part of the
faint Coma galaxies could have been recently injected from the field
following the NGC~4911 group infall. We present a list of five
Ultra Compact Dwarf galaxy candidates which need to be confirmed with
high spatial resolution imaging with the HST.  We also globally confirm
spectroscopically our previous results concerning the galaxy luminosity
functions based on imaging down to R=23 (${\rm M_R}=-12$) and find that
dwarf galaxies follow a red sequence similar to that drawn by bright
Coma galaxies.  }
{Spectroscopy of faint galaxies in Coma confirms that dwarf galaxies are
very abundant in this cluster, and that they are partly field galaxies
that have fallen onto the cluster along cosmological filaments.}

\keywords{galaxies: clusters: individual (Coma)}

\maketitle

\section{Introduction}\label{sec:intro}

On the pathway toward the use of galaxy clusters to constrain cosmology,
one must understand how clusters and their galaxy populations evolve.
Until very recently, galaxy evolution in clusters was well constrained
only down to relatively bright magnitudes (R$\leq$20 for z$\sim$0
clusters). However, according to Cold Dark Matter models of hierarchical
structure formation (e.g. White $\&$ Rees 1978, White $\&$ Frenk 1991),
there should be abundant low-mass dark-matter dominated halos present in the
Universe and these halos should therefore contain low luminosity galaxies. It is
therefore important to sample the faint and very faint cluster galaxy
populations.  Moreover, in galaxy clusters, these faint galaxies are of
major interest as their evolutionary paths are sometimes different from
those of bright galaxies: they are very sensitive to environmental
effects and can be created via interactions of larger galaxies
(e.g. Bournaud et al.  2003). They also keep their dynamical memory
longer (e.g. Sarazin 1986), and as a consequence their spatial
distribution can possibly be different from that of bright galaxies
(e.g.  Biviano et al. 1996).  With the arrival of large field cameras on
medium size telescopes, we started to reach the faint and very faint
galaxy regime. Our team concentrated on the Coma cluster
(Adami et al. 2005a,b, 2006a,b, 2007a,b, 2008, 2009a,b, Gavazzi et
al. 2009, and http://cencosw.oamp.fr/COMA/).  This cluster is relatively
nearby and this facilitated early searches (e.g. Wolf 1901, see Biviano 1998
and reference therein, and for
recent studies, see e.g. Andreon $\&$ Cuillandre 2002, Beijersbergen et
al. 2002, 
Iglesias-P\'aramo et al. 2003, Jenkins et al. 2007, Lobo et al. 1997,
Milne et al. 2007, Smith et al. 2008, Terlevich et al. 2001, Trentham
1998).  Using CFH12K and Megacam data we now have a pretty good
statistical view of the faintest galaxies existing in the Coma cluster
(M$_R \sim -9.5$).  However, this view is ``only'' statistical and we 
only have a rough idea of the behaviour of individual galaxies at these
magnitudes. Our previous spectroscopic catalog limit (despite the fact
that it gathered most of the literature data available at that time, see
Adami et al. 2005a) was far too bright even to begin investigating these
faint populations. Besides, the precision on photometric redshifts is
far too low to allow any dynamical analysis or spectroscopic
characterisation.

In order to fill this gap, we have obtained deep (21$\leq$R$\leq$23)
VIMOS/VLT spectroscopy and performed the spectroscopic characterisation of
these faint Coma cluster galaxy populations.

We describe our new spectroscopic and literature data in Sections~2 and
3.  We then present in Section~4 the analysis of the Coma line of sight
in terms of detected groups and substructures. We derive the dynamical
behaviour of the Coma cluster galaxies in Section~5 and describe the
spectral characteristics of the Coma cluster galaxies in Section~6.  In
Section~7, we build an Ultra Compact Dwarf galaxy candidate catalog. We
discuss in Section~8 the luminosity function and
color-magnitude-relation of the Coma cluster galaxies. Finally, we
summarize our results in Section~9, giving a comprehensive picture of
the Coma cluster.

In this paper we assume H$_0$ = 70 km s$^{-1}$ Mpc$^{-1}$, $\Omega
_m$=0.3, $\Omega _{\Lambda}$=0.7, a distance to Coma of 100 Mpc, a
distance modulus of 35.00, and a scale of 0.47 kpc arcsec$^{-1}$. All
  coordinates are given at the J2000 equinox.

\section{VIMOS spectroscopy}

\subsection{Settings}

Selecting targets on the basis of new photometric redshifts computed
using deep u*BVRI images, we observed three VIMOS fields in the Coma
cluster in order to spectroscopically characterize the faint cluster
population and to sample the cluster at unprecedented magnitudes of
R$\sim$23.

We obtained $\sim$1000 spectra of faint Coma line of sight galaxies
(R$\sim$[21,23]) using the VLT/VIMOS instrument in 2008 with exposure
times of $\sim$2~hours, split into five $\sim$24 minutes individual
exposures. Despite the very unfavourable declination of Coma at the VLT
latitude, we were able to observe three masks at airmasses close to 1.7,
with a seeing of the order of 1.2~arcsec (the relatively long exposure
times for R$\sim$23 galaxies compensating for the high airmass). The targets
were partly selected on a photometric redshift basis (following Adami
et al. 2008) and partly randomly in order to increase the number of
targets.

Since our aim was to obtain low resolution spectra of very faint nearby
galaxies, we used the LR-Blue grism (5.3~\AA/pix), providing a S/N
between 2 and 5, depending on the galaxy characteristics. Given the
redshift range of interest (Coma is at z$\sim$0.023), it allowed us to
efficiently sample emission lines from [OII]3727 to H$\alpha$.
 
We chose to observe strategically located regions of the Coma cluster 
(see Fig.~\ref{fig:present1}),
where infalling material has been detected (as described in Adami et al.
2005a): the west infalling galaxy layer and substructures close to
NGC~4911.

We obtained a reliable redshift (reliability flag $\geq$ 2, see the 
following)
for 715 objects. Among these, slightly less than 100 galaxies are part
of the Coma cluster. The minimum number of galaxies expected inside the
Coma cluster given the target selection was 70, so our results are in
good agreement with our predictions.

\subsection{Data reduction}

The spectra were extracted using the VIPGI package (Scodeggio et
al. 2005), which was extensively used for the VIMOS VLT Deep Survey
(VVDS) data (e.g. Le F\`evre et al. 2005). This package (VIMOS
Interactive Pipeline and Graphical Interface: VIPGI), is a new software
designed to simplify to a very high degree the task of reducing
astronomical data obtained with VIMOS (Visible MultiObject
Spectrograph). The final product was the wavelength-calibrated 1D spectra of
our targets.

Due to the fact that our observations were made at high airmasses,
  differential refraction was not negligible. However, the level of
  spectra physical curvature on the CCDs remained modest and was easily
  compensated by a linear function to define the extraction area. The
  slope of the linear function was visually adjusted for each
  quadrant.

We then adopted the same strategy as for the VVDS. Each spectrum was examined
independently by two of us (CA and VLB) and redshifts were
proposed using well known cross-correlation techniques such as the $rvsao$
IRAF package or the $EZ$ VVDS tool (as well as VIPGI-included line gaussian 
fitting when possible).  Finally, the different redshift determinations were 
compared and a final list was established, including quality flags.

When a redshift determination was possible, it was flagged with a number
between 1 and 9, following the VVDS conventions (e.g. Le F\`evre et
al. 2004) and indicating the reliability level of the measurement. Flags
2, 3, 4 are the most secure, flag 1 is an indicative measurement based
on few supporting features, and flag 9 indicates that there is only one
secure emission line corresponding to the listed redshift. 
Le F\`evre et al. (2005) have shown with repeated observations that these
flags correspond to the following probabilities of finding the correct
redshift: 55$\%$ for flag 1, 81$\%$ for flag 2, 97$\%$ for flag 3, and
99.5$\%$ for flag 4. Given these numbers we simply chose to ignore all
objects with flag 1.

We also gave a second integer flag between 0 and 3 (spectral flag): 0
for absorption line only spectra, 1 for spectra showing both emission
and absorption lines, 2 for spectra showing only emission lines, and 3
for active objects (i.e. showing broad band emission lines or very
strong Balmer lines as compared to oxygen lines).

We finally note that our spectra are not photometrically calibrated.

\subsection{VIMOS spectroscopic sample}

Table~\ref{centerf} gives the basic details of the spectroscopic VIMOS observations.
Out of the 926 slits, 833 objects provided a spectrum with a tentative
redshift measure. Among these 833, 118 were flagged with a reliability flag of
1, 298 with a reliability flag of 2, 187 with a reliability flag of 3, 187
with a reliability flag of 4, and 43 with a reliability flag of 9.

After excluding objects with a reliability flag equal to 1, we were left
with 715 galaxies, out of which 250 had a spectral flag of 0, 288 a
spectral flag of 1, 168 a spectral flag of 2, and 9 a spectral flag of
3.

Figs.~\ref{fig:present2}, ~\ref{fig:present3}, and
~\ref{fig:present4}  give a general idea of the properties of the sample:
distribution of the redshifts as a function of the R-band magnitudes, redshift 
histogram, and magnitude histogram.

\begin{table}
\caption{For the three VIMOS fields observed: coordinates (J2000),
number of slits (N) and exposure times.}
\begin{center}
\begin{tabular}{ccccc}
\hline
field number & $\alpha$   &  $\delta$ & N & Exp. time \\
             & (deg)   & (deg) &   & (s)  \\
\hline
1 &   194.72     &   27.87              & 322 &  7200  \\ 
2 &   194.69     &   28.14              & 298 &  7184  \\ 
3 &   195.04     &   27.72              & 306 &  7098  \\ 
\hline
\end{tabular}
\label{centerf}
\end{center}
\end{table}

\begin{figure}
\caption[]{u* band Megacam image with area covered by the VIMOS
spectroscopy overlayed (in red). Large red circles are the VIMOS
galaxies inside the Coma cluster. Small green circles are galaxies
inside the Coma cluster taken from the literature. Blue contours
represent the X-ray substructures from Neumann et al. (2003). Coordinates are J2000.}
\label{fig:present1}
\end{figure}

\begin{figure}
\centering \mbox{\psfig{figure=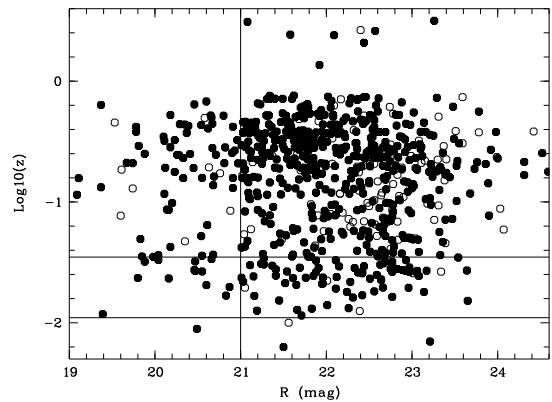,width=8cm,angle=270}}
\caption[]{Logarithm of the redshift versus Vega R band magnitude for
the spectroscopic sample. Filled circles are galaxies with a reliability
flag at least equal to 2. Open circles are galaxies with a reliability
flag equal to 1. The two horizontal lines show the redshift limits
chosen in this paper for the Coma cluster. The vertical line is the
minimum R band magnitude we will consider for the most of the analyses (R=21).}
\label{fig:present2}
\end{figure}

\begin{figure}
\centering \mbox{\psfig{figure=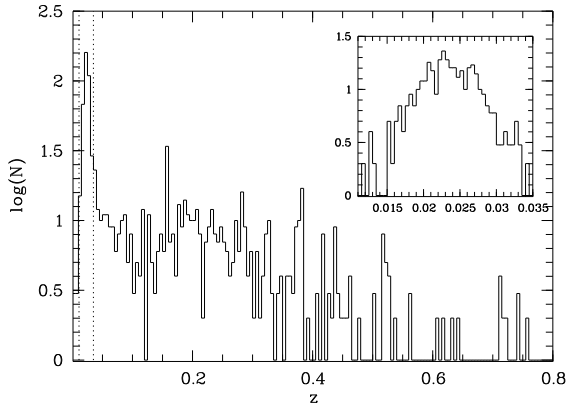,width=8cm,angle=270}}
\caption[]{Redshift histogram of the spectroscopic data (bin of 0.005) 
available in the
region of interest (literature + VIMOS). The two vertical dotted lines
represent the limits we adopted for the Coma cluster. The inner box shows the
redshift histogram (bin of 0.0005) for the Coma cluster itself.}  
\label{fig:present3}
\end{figure}

\begin{figure}
\centering \mbox{\psfig{figure=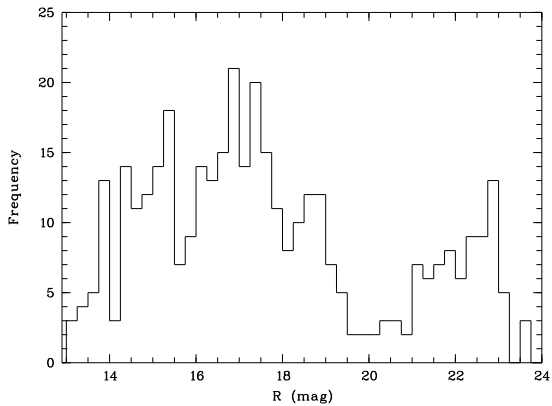,width=8cm,angle=270}}
\caption[]{R band magnitude histogram of the spectroscopic data
available in the region of interest (literature + VIMOS) for the
galaxies inside the Coma cluster.}  \label{fig:present4}
\end{figure}

We also investigated the velocity uncertainty resulting from our
measurements. With a similar instrumental configuration, Le F\`evre et
al. (2005) estimated the redshift uncertainty to be of the order of 280
km/s from repeated VVDS redshift measurements. We did the same exercise
by considering the literature redshifts included in our VIMOS sample
(6 galaxies between R=17 and 19.8) or observed twice during our
spectroscopic run (3 galaxies between R=20.8 and 23), and with a
reliability flag strictly greater than 1. Given the reliability flags,
we expect to have 8 galaxies over 9 providing a good agreement.  In
practice, 7 show a good agreement. One galaxy has a 0.008
difference in redshift, but checking the literature spectrum (from the
SDSS survey) and applying our redshift measurement methods, we estimate
a redshift similar to the one we obtained from our own spectrum of the
same galaxy (the SDSS automated measure pipeline misinterpreted the edge
of an absorption line as an emission line).

\begin{figure}
\centering \mbox{\psfig{figure=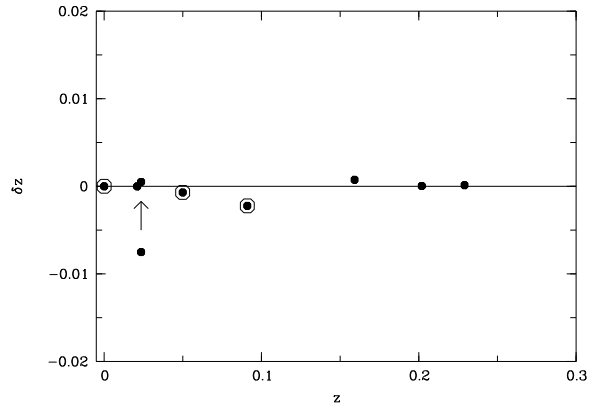,width=8cm,angle=270}}
\caption[]{Difference between literature and VIMOS redshift measurements
as a function of redshift. The galaxy with the 0.008 difference (see
text) is shown before and after remeasuring its redshift.The circled
  objects are the three galaxies observed two times in our spectroscopic runs
  and are fainter than R=20.8 (see text). }
\label{fig:comparz}
\end{figure}

For another galaxy there is a strong redshift discrepancy between our
measurement and that of the literature. However, our spectrum
corresponds to a flag 4 (z=0.1468) galaxy with strong emission lines,
whereas the spectrum from the literature is not available for
inspection, making our present determination more reliable than the
previous one.  Fig.~\ref{fig:comparz} shows the resulting redshift
comparison, corresponding to a statistical uncertainty of 196~km/s,
slightly lower than the VVDS estimate. We also show in this graph
the galaxies observed twice in our spectroscopic runs. Despite the low
statistics, these objects do not seem to show a different behaviour compared to
brighter objects in the literature.

\section{Complementary data}

\subsection{Spectroscopic data from the literature}

A spectroscopic catalog based on data taken from the literature was
compiled by one of us (RT) from the NED and SDSS (including up to DR6) 
databases. We extracted from this catalog all galaxies in our
VIMOS region which are not already present in our VIMOS data. We also
added a few new redshifts from the SDSS DR7. The resulting literature
catalog is therefore more complete than the one we used in Adami et
al. (2005a). These additional data were included in Figs.~\ref{fig:present2}, 
\ref{fig:present3}, and~\ref{fig:present4}. Fig.~3 shows the Coma cluster 
itself plus several
structures along the line of sight for example at z$\sim$0.16, 0.38, or
0.52. Fig.~4 shows the magnitude gap between spectroscopic data from the
literature and our new VIMOS spectroscopic data. This gap has no
consequences on most of the following analyses which are based on R$\geq$21
objects. We have however to bear in mind
  the magnitude distribution of the spectroscopic sample when discussing about
  possible structures along the line of sight.

\subsection{u*, B, V, R and I band CFHT imaging data}

The B, V, R, and I data are fully described in Adami et al. (2006a).  We
give here only the salient points. A mosaic of two fields was observed
with the CFH12K camera in four bands (B, V, R and I) covering a
52$\times$42~arcmin$^2$ field. The total field is approximately centered
on the two dominant cluster galaxies (NGC~4874 and NGC~4889). We derived
total Kron Vega magnitudes (Kron 1980) from these images. The
completeness level in R is close to R$\sim$24. The seeing conditions
were all close to 1 arcsec. All these imaging data are available at
http://cencosw.oamp.fr.

The u* data including the previous field are described in Adami et
al. (2008). They were obtained between 2006 and 2007 with the CFH
Megacam camera in a field of view of 1 deg$^2$ with an average seeing of 
1.1~arcsec. The total exposure time was 9.66~hours.  We derived total
Kron AB magnitudes from this image. Fig.~\ref{fig:present2} shows a
subarea of the total u* image.

These data therefore provide a catalog of objects in the u*, B, V, R and
I bands complete down to R$\sim$24.

\section{Groups along the Coma cluster line of sight from the spectroscopic sample}

We now present our analysis of the groups found along the Coma line of
sight.

First, we have to define the boundaries of the Coma cluster in terms of
velocity. From Adami et al. (1998), we know that the Coma cluster galaxy
velocity dispersion $\sigma _v$ increases with decreasing luminosity, as
expected from dynamical considerations. A natural way to fix cluster
boundaries is to limit the cluster velocity range within $\pm 3\sigma
_v$. Assuming the mean $\sigma _v$ for the faintest galaxies in Adami et
al. (1998) (1200 km/s for R$\sim$17.5), we limit the cluster to the
redshift range [0.011;0.035].

Second, in order to determine galaxy orbits, we limit the spectroscopic
sample to the areas where photometric redshifts are also available (in order
to be able to compute a density profile on the photometric redshift basis).
Finally we restrict the magnitude range of the spectroscopic sample to
the R=[21,23] range where the sampling rate is the most homogeneous.

In order to search for galaxy groups along the Coma cluster line of
sight, we applied the Serna-Gerbal (SG hereafter: Serna $\&$ Gerbal
1996) method to our total redshift catalogue (VIMOS + literature). This
hierarchical method was already applied in one of our early studies
(Adami et al. 2005a) and we refer the reader to this paper for more
details. Briefly, it allows galaxy subgroups to be extracted from a
catalogue containing positions, magnitudes, and redshifts, based on the
calculation of their relative (negative) binding energies. Note that
this calculation takes into account the mass to luminosity $M/L$ ratio
chosen by the user as an a priori input value of the $M/L$ ratio in
the structure. The group mass derived later is estimated from the group
binding energy and velocity dispersion, and does not depend upon M/L,
which proves empirically (e.g. Covone et al. 2006) to act mainly as a
contrast criterion. Results do not depend strongly on this factor,
since a variation of a factor of two in this parameter does not change
significantly the results. Qualitatively, low input values of M/L allow
us to detect structures with low binding energies, while high values of
M/L only allow the detection of the major structures.

The output of the SG method is a list of galaxies belonging to each group,
as well as information on the binding energy and mass of the group itself. We
will consider here that a group consists of at least 3 members.

We use a nominal M/L ratio in the R band of 400, and
we also search for less strongly linked galaxy groups inside the Coma
cluster and inside the z$\sim$0.5 large scale structure (see the following)
assuming lower M/L ratios of 100.

\begin{figure}
\centering \mbox{\psfig{figure=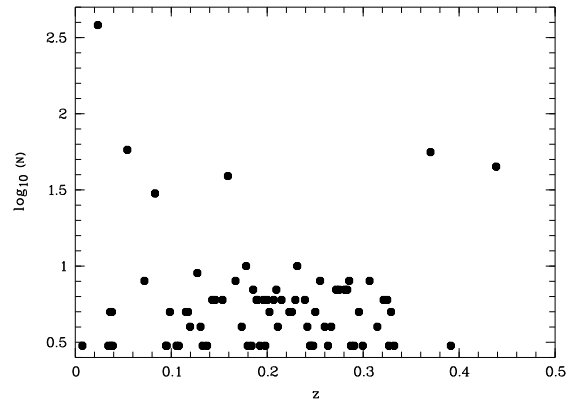,width=8cm,angle=270}}
\caption[]{Numbers of group members found by the Serna--Gerbal method
 (in logarithmic scale) versus redshift for the 76 N$\geq$3 detected
groups.}  \label{fig:SG1}
\end{figure}

The SG method reveals the existence of 76 ``groups'' along the line of
sight (assuming M/L=400), including the Coma cluster itself. Among them,
52 groups have $N\ge4$ members, and 44 groups have $N\ge5$ members.
In Fig.~\ref{fig:SG1} we show the log$_{10}$(N) versus redshift for the
76 groups. There is no relation between N and the group redshifts.

The main caveat of this analysis is that we do not have a complete
spectroscopic catalog. We will dedicate the following section to this
point.

\subsection{Sampling of the total spectroscopic catalog}

In order to find all the structures along the Coma cluster line of
sight, the SG method would require the galaxy catalogue to be 100$\%$
complete, which is obviously never the case. The detection efficiency
depends on the completeness of the spectroscopic catalog and we must
therefore analyze this completeness in the zone that we considered. For
this, we use our CFHT R-band photometric catalog
(http://cencosw.oamp.fr/).  We pixelize the VIMOS field in several
subregions of 0.2$\times$0.2~deg$^2$ and compute the percentage of
objects with a redshift available in our spectroscopic catalog (VIMOS +
literature), as a function of R band magnitude.  Fig.~\ref{fig:comp}
gives the sampling rate in the VIMOS area for several magnitude ranges.

\begin{figure*}
\centering
\caption[]{Completeness level in percentage in the VIMOS area (red
limited regions) in the R magnitude ranges [18,20] (upper left), [19,21]
(upper right), [20,22] (lower left), and [21,23] (lower right).
Coordinates are J2000.}  \label{fig:comp}
\end{figure*}

\subsection{Substructures in the Coma cluster}

Assuming an M/L of 400, well adapted to find highly gravitationally bound
structures, we detect the Coma cluster as a single structure (z$\sim$0.0234
sampled with 382 redshifts), along with a minor additional substructure
(sampled in spectroscopy by only 3 galaxies). 

We then investigate how the main detected structure can be split in
several substructures. For this, we redo the same SG analysis using this time a
M/L ratio of 100. Results are shown in Fig.~\ref{fig:SG2}. We
essentially confirm in the considered area the results of Adami et
al. (2005a), and detect 7 subtructures.  Three are directly linked with
NGC~4874 (G1 of Adami et al. 2005a). A fourth one (undetected in Adami et
al.  2005a) is detected north of NGC~4874 and coincides with the
northern part of the west X-ray substructures of Neumann et al. (2003).
The last three are detected south of NGC~4874. One of them is along the
NGC~4839 infalling path, and coincides with the G8/G9 groups (see
Table~2 and Fig.~4 in Adami et al. 2005a), while the last two were not
detected in Adami et al. (2005a). These three groups also seem to be
visible in the Okabe et al. weak lensing study (private communication)
and are included in the south-west mass extension detected by Gavazzi et
al. (2009), who used the same photometric data as in the present paper.

\subsection{Structures at z$\leq$0.2 not belonging to Coma}

Besides the Coma cluster substructures, we are more
specifically interested in the z$\leq$0.2 structures because they cannot
be efficiently removed from the background using only photometric
redshift techniques (see Adami et al. 2008).

We give in Fig.~\ref{fig:SG2} and in Table~\ref{b02} the locations and
characteristics of the z$\leq$0.2 structures, excluding the structures
which are members of the Coma cluster.

\begin{figure}
\caption[]{u* band image overlayed with X-ray substructures from Neumann
et al. (2003: blue contours), and with the mass map from Gavazzi et
al. (2009: green contours), VIMOS fields (red area), and galaxy groups
at z$\leq$0.2 and outside the Coma cluster (red: z$\geq$0.1, green:
z$\leq$0.1).  Coma substructures from the M/L=100 SG analysis are shown
as black circles. PFA is located around $\alpha$=194.77~deg and
$\delta$=27.91~deg. Coordinates are J2000.}  \label{fig:SG2}
\end{figure}

\begin{table}
\caption{Characteristics of the structures at z$\leq$0.2 not belonging
to Coma (in order of increasing redshift).}
\begin{center}
\begin{tabular}{ccccc}
\hline
$\alpha$   &  $\delta$ & N & Tentative mass & redshift \\
 (2000.0)  &  (2000.0) &   & (if virialized) & \\
  (deg)    &  (deg)    &   &      (M$\odot$)          & \\
\hline
   194.9867     &     27.77410     &  	        3    &    6.67E+13   &  0.0074   \\ 
   194.7974     &     27.93770      &          5      &   3.9E+12   &  0.0364    \\
   194.8937     &     27.81673     &           3     &    2.0E+12   &  0.0373    \\
   194.7462     &     27.87338      &          5     &    2E+10   &  0.0381    \\
   194.8206     &     27.95107      &         58     &    2.73E+15   &  0.0542   \\ 
   194.8296     &     27.84444     &           8     &    1.27E+14   &  0.0720   \\ 
   194.8266     &     27.95877     &          30    &     1.88E+14   &  0.0831   \\ 
   194.8543     &     27.92563     &           3     &    1.8E+12  &   0.0945   \\ 
   194.7782     &     28.16868     &           5     &    7E+11   &  0.0986  \\ 
   194.7517     &     28.17983     &  	        3    &    5E+11  &   0.0952  \\ 
   194.9777     &     27.78710     &           3     &    7.2E+12  &   0.1054        \\
   194.8250     &     27.90883      & 	        3    &    6E+11  &   0.1078        \\
   194.7706     &     27.82888     & 	        5    &    8.8E+12  &   0.1153       \\ 
   194.7916     &     27.89518     &  	        5     &   5.3E+12  &   0.1175       \\ 
   194.7713     &     27.94355     &           4      &   1.6E+12  &   0.1197       \\ 
   194.8791     &     27.85061      &          9     &    1.19E+13   &  0.1272      \\  
   194.8550     &     27.92140      &          4     &    1.86E+13  &   0.1305       \\ 
   194.6917     &     27.89147      & 	        3    &    1E+08  &   0.1326       \\ 
   194.8280     &     27.82573      &          3     &    4.18E+13  &   0.1374      \\  
   194.8205     &     28.02470      &          6     &    8.23E+13  &   0.1427       \\ 
   195.0567     &     27.73113      &          6     &    1.9E+12   &  0.1458       \\ 
   194.7197    &      28.09023     &  	        6    &    3.7E+12  &   0.1534        \\
   194.7230     &     27.89113    &           39     &    8.5E+12  &   0.1590        \\
   194.7263    &      27.99955     &           4     &    8.13E+13  &   0.1734        \\
   194.9545    &      27.85410      &         10     &    2.7E+12  &   0.1780       \\ 
   194.8757    &      27.91133     &           3     &    1.1E+12  &   0.1796       \\ 
   195.0393    &      27.76123      &          3     &    3E+10  &   0.1838        \\
   194.9579    &      27.83376      &          7     &    4.68E+13  &   0.1854       \\ 
   194.8118     &     27.89033     &           6     &    3E+11  &   0.1888       \\ 
   194.8523     &     28.01927      &          6     &    2.3E+12  &   0.1900       \\ 
   194.6847     &     27.93407     &           3     &    8E+11  &   0.1925       \\ 
\hline
\end{tabular}
\label{b02}
\end{center}
\end{table}

\begin{table}
\caption{Characteristics of the substructures  detected in the Coma cluster.}
\begin{center}
\begin{tabular}{ccccc}
\hline
$\alpha$   &  $\delta$ & N & Tentative mass & redshift \\
 (2000.0)  &  (2000.0) &   & (if virialized) & \\
  (deg)    &  (deg)    &   &      (M$\odot$)          & \\
\hline
194.9353        &   27.83393       &  3	          &  3.85E+11     &   0.0117  \\ 
194.8504        &   27.96517       &  7	          &  4.83E+10     &   0.0128  \\ 
194.8732        &   27.95258       &  9	          &  2.83E+12     &   0.0162  \\ 
194.6830        &   27.84417       &  4	          &  2.49E+12     &   0.0193  \\ 
194.8887        &   27.95417       &  3	          &  1.17E+12     &   0.0236  \\ 
194.7610        &   28.15490       &  3	          &  1.36E+13     &   0.0283  \\ 
194.8395        &   27.82490       &  4	          &  5.08E+12     &   0.0342  \\ 
\hline
\end{tabular}
\label{comass}
\end{center}
\end{table}

We immediately notice a concentration of structures in a small zone
south-west of NGC~4874, which we will call the Putative Filament Area
(PFA) in the following, while other parts of the spectroscopic area only present a sparse
distribution of z$\leq$0.2 background groups. The location of these detections
cannot be explained by the spectroscopic incompleteness (see
Fig.~\ref{fig:comp}) alone. Even if not uniform, this sampling is not
preferentially high in the PFA.

Part of the detected groups are included in the SDSS Great Wall (Gott et
al. 2005) and are coincident with the PFA, confirming the existence of a complex structure between
z$\sim$0.07 and 0.085.  We also detect a new large galaxy structure at
z$\sim$0.054 (the Background Massive Group, or BMG hereafter). This
structure, undetected with SDSS data alone, is mainly populated by faint
galaxies and is very rich (sampled with 58 spectroscopic redshifts). It
is probably not virialized because applying the virial theorem would
lead to a mass of $\sim$2.7 10$^{15}$M$_\odot$. Such a major mass
concentration would evidently show up in X-rays, and nothing is detected
by Neumann et al. (2003) at this location. 

This does not mean that this structure has a negligible mass however,
and it could significantly contribute to the mass concentration detected
with a weak lensing analysis by Gavazzi et al. (2009) at the same
location (see Fig.~\ref{fig:SG2}). If we define the PFA as centered at
$\alpha$=194.772, $\delta$=27.914 and enclosed in a circle of 5.5'
radius, we can estimate a mass from the Gavazzi et al. (2009) weak
lensing analysis of (4.2 $\pm$ 0.8) 10$^{13}$~M$_\odot$ (if at z=0.023),
or 2~10$^{13}$ M$_\odot$ (if at z=0.055). The Adami et al. (2005a) G8
and G9 Coma substructures are probably contributing to the PFA mass
detected by weak lensing. However, G8 and G9 are not detected in X-rays
and their maximum mass can be inferred to be $\sim$0.5 10$^{13}$
M$_\odot$, the lowest mass Coma substructure emitting in X-rays (the
group attached to NGC~4911, see Neumann et al. 2003). Groups G8 and G9
can therefore account for $\sim$25$\%$ of the PFA estimated mass. The
75\% remaining could then come from the z$\sim$0.054 BMG structure,
leading to a mass of the order of $(1-3)\ 10^{13}$~M$_\odot$. This mass
is high enough to justify the qualification of massive, but small enough
to justify the undetected X-ray emission. The large difference of this
mass estimate with the virial one clearly argues in favor of an
unvirialized structure.  With a crossing time of $\sim$2.9 10$^{8}$
years at z=0.054, this places the BMG structure in a very early
evolutionary stage.

Several other smaller structures are detected
between z=0.035 and 0.2 in the PFA. This leads us to suspect
the existence of a (minor) line of sight filament joining the Coma
cluster and the new z$\sim$0.054 large galaxy structure, and then
extending toward the SDSS Great Wall and beyond.

Such a high level of structures in the immediate background vicinity of
the Coma cluster could have a significant effect on the cluster
luminosity function determinations in this region, estimated for example
from statistical arguments (see Adami et al. 2007a,b), because a
significant part of these nearby background groups were not detected in
this early study. We will therefore determine in the following a
luminosity function only based on our spectroscopic data.

We note that most of these structures (e.g. the BMG or the SDSS Great
Wall) do not prominently appear in Fig.~\ref{fig:present3} because they
are not fully virialized and therefore their redshift distribution is
not very compact.

\subsection{Sampling the Coma back-infalling galaxy layers: nature of galaxy haloes}

\begin{figure}
\centering \mbox{\psfig{figure=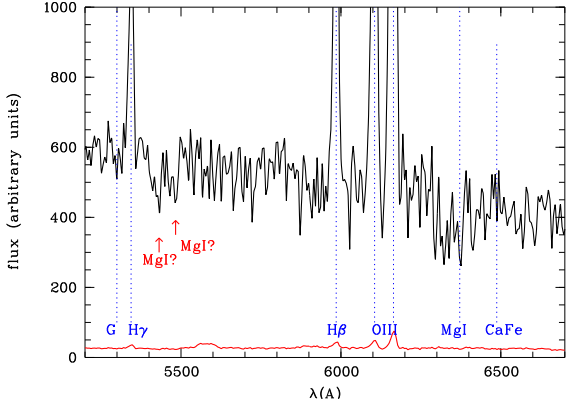,width=8cm,angle=270}}
\caption[]{VIMOS spectrum of a z=0.2312 active object (black) detected in our
survey. The noise spectrum is shown in red. The possible foreground MgI
lines at z=0.0496 and z=0.0595 are indicated in red.}  \label{fig:sdss}
\end{figure}

Our spectroscopic sample contains a number of active objects, which can
be used to study the foreground gaseous clouds through the absorption
features that imprint the spectrum (e.g. Ledoux et al. 1999). This method has 
led to the discovery
of extended halos around field galaxies (e.g. Bergeron $\&$ Boiss\'e
1991) and of the intergalactic medium (the well known Lyman-alpha
forest, see e.g. Croft et al. 2002, for a detailed review), and has
allowed to study the internal regions of high redshift galaxies (through
the so-called Damped Lyman-alpha systems, see Khare et al. 2007). We
tried to detect the gaseous regions just behind the Coma cluster with
the same method. However the case is particular, because of the very low
redshift of Coma and of the wavelength range of our spectroscopic
observations, which impedes us from using high-redshift background
targets, since it would have been impossible to disentangle blue
restframe wavelength lines at high redshift from redder restframe
wavelength lines at low redshift. We therefore looked for low redshift
(z$\leq$0.25) active objects in our spectroscopic data. At those redshifts, the only strong absorption
lines that could arise from gaseous halos or interstellar medium, and be
detectable in our spectra are MgI and Ca H\&K. Other usual absorption
lines like MgII or CIV have rest frame wavelengths in the UV, and are
not detectable in the optical part of a low redshift spectrum. 
One object is suitable for our study at a redhift of 0.2312 and
located spatially close to the BMG. The
spectrum of this VIMOS object is displayed in Fig.~\ref{fig:sdss} and is
probably a Seyfert 2 galaxy.

As can be seen on the magnified part of the VIMOS spectrum, there is an
unidentified double-line, shortward of the intrinsic MgI line. The respective
equivalent widths are 2.8 and 2.5~\AA, and the detection levels are
$3\sigma$. The absorption redshifts correspond to z$\sim$0.0496 and
0.0595, just behind the Coma cluster and very close to the mean redshift
of the BMG structure discussed in Section~4.3. The only possible
identification is MgI, as any other identification would either lead to
a negative redshift or to a redshift higher than that of the active
object.  However, MgI is a low ionization line, which arises only in low
temperature haloes around field galaxies, and definitely not in high
temperature intracluster gas. This therefore suggests that the BMG
intrastructure medium does not have a high temperature, consistent with
its supposedly unvirialized dynamical status.

We have searched for galaxies located within 100~kpc of the line of
sight to the VIMOS active object, as this corresponds to the measured
size of low ionization haloes around field galaxies
(Fig.~\ref{fig:vimosi}). There is a galaxy at redshift 0.0482, located
$\sim 50 h^{-1}$~kpc away from the active object, plus several galaxies
potentially at z$\leq$0.2 (from photometric redshift estimates by Adami
et al. 2008). Given that probably most field galaxies are surrounded by
a gaseous halo of radius $\sim$90~kpc (Kacprzak et al. 2007) we suggest
that one or more of these galaxies (part of the BMG structure) also has
its own halo. The intrastructure medium of the BMG therefore allows a gaseous 
halo to survive in at least
one of these galaxies. This is in good agreement with the early
evolutionary stage of the BMG.

\begin{figure}
\caption[]{Vicinity of the z=0.2312 VIMOS active object. The large blue
circle shows a 100~kpc radius at z=0.055. The small blue circle shows
the position of the active object. The green circle is the only galaxy
with a known spectroscopic redshift in the field. Red circles are
galaxies at photometric redshifts z$\le$0.2. Coordinates are J2000.}  \label{fig:vimosi}
\end{figure}

\subsection{Structures at z$\sim$0.5 }

In order to characterize the vicinity of the z$\sim$0.5 galaxy structure
detected in Adami et al. (1998 and 2000), we are also interested by the
structures near that redshift. Assuming an M/L ratio of 100, we find with the
SG method a very extended structure (covering the whole field of view)
at z$\sim$0.52 and sampled with 29 spectroscopic redshifts
(Fig.~\ref{fig:SG05}), which we do not detect with an M/L ratio of
400. A more concentrated structure is found inside this galaxy layer
(sampled with 6 galaxies), located on top of the Adami et al. (1998 and
2000) cluster candidate at z$\sim$0.52. We therefore have a
z$\sim$0.52 compact structure of galaxies in this zone. We cannot
estimate a robust mass because of the low sampling. However, the fact
that we detect this structure neither in the weak lensing mass map of
Gavazzi et al. (2009) nor in the X-ray maps of Neumann et al. (2003) and
Forman et al. (private communication) does not speak in favor of a very
massive structure.

\begin{figure}
\caption[]{u* band image with the VIMOS area (red area) with the
z$\sim$0.52 galaxy structure members (green circles) overlayed. Coordinates are J2000.}
\label{fig:SG05}
\end{figure}

\section{Galaxy orbits}

It is important to characterize the orbits of faint galaxies in Coma as
this is a powerful tool to put constraints on their origin. While bright
galaxy orbits are relatively well known in clusters (e.g. Biviano $\&$
Katgert 2004), nothing is known about the orbits of faint dwarf
galaxies.  We will select in the following all galaxies with magnitudes
between R=21 and 23 and having a measured spectroscopic redshift . We
assume as a first guess that the dwarf galaxy population is homogeneous
in order to perform the Jeans analysis.

\subsection{Jeans analysis}

In order to determine the orbits of a population of galaxies in a cluster, we
need three ingredients:

\begin{itemize}
\item the mass profile of the cluster, $M(r)$, where $r$ is the 3-D
distance from the cluster center;
\item the projected number density profile of the galaxies, $N(R)$, 
where $R$ is the projected distance from the cluster center;
\item the line-of-sight velocity dispersion profile of the 
galaxies $\sigma(R)$. 
\end{itemize}

We choose to take $M(r)$ from the literature, scaling all parameters to
the presently assumed cosmology. Geller et al. (1999) have applied the
caustic technique to determine the mass profile of the Coma cluster out
to large radii. This technique is claimed to provide a cluster mass
profile with little dependence on the assumed orbital velocity
anisotropy. They have fit the mass profile with a NFW model with
concentration $c=r_{200}/r_s=8.3_{-1.2}^{+1.7}$ and mass $M_{200}=1.14
\times 10^{15} \, M_{\odot}$. Another study by {\L}okas \& Mamon (2003)
gives for the dark matter profile $c=9$ and a total mass at the virial
radius of $M_{200}=1.19 \times 10^{15} \, M_{\odot}$. These are not very
different from the Geller et al. estimates, and support our initial
choice.

In both Geller et al.'s and {\L}okas \& Mamon's analyses, the Coma
cluster center is taken to be the position of NGC~4874. Since we base
our analysis on their mass profile, we must consistently assume the same
center throughout our analysis (i.e. also for $N(R)$ and $\sigma(R)$).

We then consider the value of $M_{200}$ given by Geller et al. to define
the scaling radius $r_{200}$ and the scaling velocity $v_{200} \equiv
(G M_{200} / r_{200})^{1/2}$, so that we can work in the space of
normalized radii $r_n \equiv r/r_{200}$ ($R_n \equiv R/r_{200}$ in
projection), and velocities $v_n \equiv v/v_{200}$, where $v$ are the
line-of-sight velocities with respect to the cluster mean velocity,
corrected for the cosmological term, $v \equiv (V_{los}-V_c)/(1+V_c/c)$,
where we have taken $V_c=7090$~km~s$^{-1}$ from Geller et al.

We have checked that if we take {\L}okas and Mamon's value for $M_{200}$
instead of Geller et al.'s, the results of the Jeans analysis are
essentially unchanged.

\begin{figure}
\begin{center}
\begin{minipage}{0.5\textwidth}
\resizebox{\hsize}{!}{\includegraphics{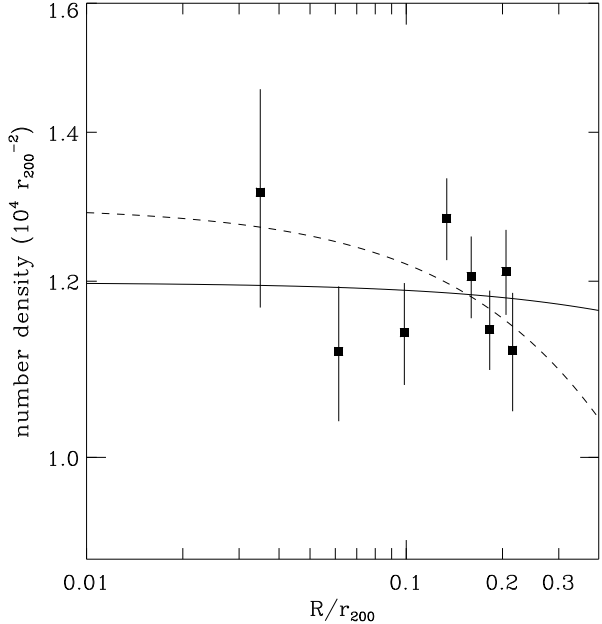}}
\end{minipage}
\end{center}
\caption{Number density profile $N(R_n)$.
~1$\sigma$ error bars are shown. The solid line is the best-fit core
profile, the dashed line is the best-fit King-model profile (i.e. a core
profile with slope $a=-1$). } \label{f-NR}
\end{figure}

In order to compute $N(R_n)$ one must be aware of the problems due to
incompleteness. One cannot use the sample of spectroscopically confirmed
members, since it is only poorly complete. We therefore consider here
the sample of dwarfs whose membership is based on their photometric
redshift $z_p<0.2$. We computed a binned $N(R_n)$ by counting galaxies
within circular annuli and dividing these counts by the effective area, i.e. 
the total area of annuli excluding the masked regions (defined in Adami et
al. 2006a).  The resulting $N(R_n)$ is shown in Fig.~\ref{f-NR}.

We fit $N(R_n)$ with a core model profile, $N(R_n)=N_0
[(1+(R_n/R_c)^2]^{-a}$, with three free parameters. Note that $N_0$ does
not enter the Jeans equation solution (it cancels away), so the
effective number of interesting parameters is two. The best fit obtained
with $N_0=11544$ $r_{200}^{-2}$, $R_c=1.2 \, r_{200}$ and $a=-0.1$ is shown
in Fig.~\ref{f-NR}. Clearly, the density of dwarf galaxies is almost
constant with $R_n$ (at least in this central region).

\begin{figure}
\begin{center}
\begin{minipage}{0.5\textwidth}
\resizebox{\hsize}{!}{\includegraphics{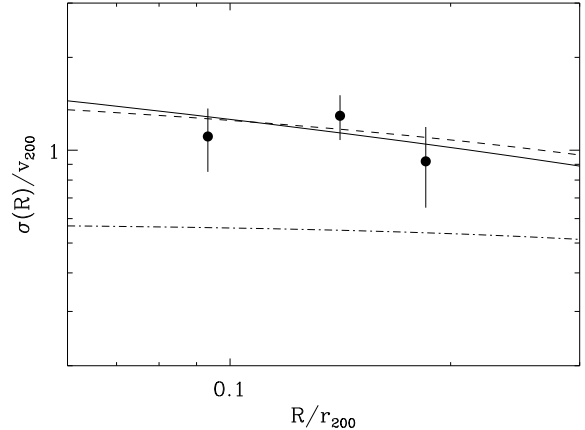}}
\end{minipage}
\end{center}
\caption{Velocity dispersion profile. 1$\sigma$ error bars are
shown. The solid line is the Jeans solution in the assumed NFW mass
profile for $\beta'=1.8$, obtained using the best-fit core profile for
$N(R_n)$. The dashed line is the Jeans solution in the assumed NFW mass
profile for $\beta'=1.3$, obtained using the best-fit King-model profile
for $N(R_n)$. The dash-dotted line is the isotropic Jeans solution
($\beta'=1$) in the assumed NFW mass profile, obtained with the best-fit
core profile for $N(R_n)$.}  \label{sigma}
\end{figure}

For the determination of $\sigma(R_n)$ we consider the spectroscopic
sample. We do not need to care about incompleteness, as long as the
velocity dispersion is independent of the galaxy magnitude. This is
likely to be the case since dwarf galaxies are very low-mass objects and
the timescale of dynamical friction needed to slow down such light
objects is beyond the Hubble time in a cluster like Coma. We then
determine $\sigma$ for the galaxies in circular annuli. The resulting
binned profile is shown in Fig.~\ref{sigma}.

\begin{figure}
\begin{center}
\begin{minipage}{0.5\textwidth}
\resizebox{\hsize}{!}{\includegraphics{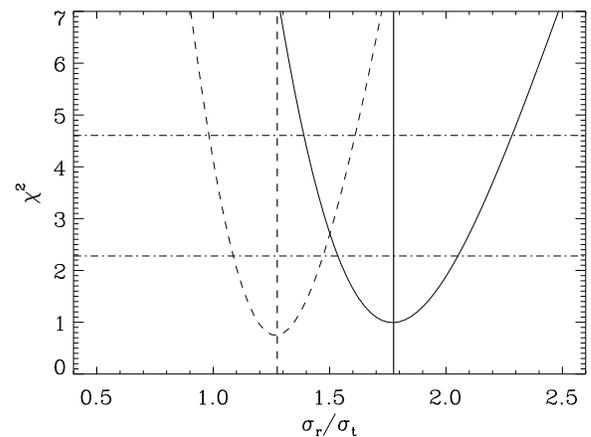}}
\end{minipage}
\end{center}
\caption{Results of the Jeans analysis: $\chi^2$ values for different
  $\beta'$ values. The solid line shows the results obtained using the
  best-fit core profile for $N(R_n)$, the dashed line shows the results
  obtained using the best-fit King-model profile for $N(R_n)$. The
  vertical lines indicate the best-fit $\beta'$ values. The horizontal
  lines represent the $\chi^2$ 68\% and 90\% limits.}  \label{chi2}
\end{figure}

\subsection{Results}

We are now in the position of solving the Jeans equation. Ideally, we
  could perform this analysis splitting our sample in several
  spectromorphological types, but the statistics are too low for this purpose. 
We note however that our sample is dominated by absorption line
galaxies (with or without emission lines) and by relatively red objects
($\sim$75$\%$ of the considered sample is redder than the 2-$\sigma$ lower bound of the
Color Magnitude Red Sequence, see Fig.~\ref{fig:CMR_RS}). Our results will
therefore mainly apply to red dwarf galaxies.

Given $M(R_n)$
and $N(R_n)$ we search for the constant value of $\beta \equiv
1-(\sigma_{t}/\sigma_{r})^2$ that provides the best-fit to
$\sigma(R_n)$, where $\sigma_t$ and $\sigma_r$ are respectively the
tangential and radial components of the velocity dispersion.  A constant
$\beta$ model should be adequate here given that the sampled region does
not extend over a wide radial range. We prefer to express the results in
terms of $\beta' \equiv \sigma_r/\sigma_t$.  The best fit is given by
$\beta'=1.8 \pm 0.3$, significantly different from the isotropic case
$\beta'=1$. This can be seen in Fig.~\ref{chi2}.

Should we force isotropy with $\beta'=0$, we would obtain an unacceptable
solution. On the other hand, should we force isotropy and let the
concentration of the mass profile free to vary, we would obtain
marginally acceptable solutions (90\% confidence level) but only for
very low concentrations, $c<0.3$, which are excluded by all analyses
of the mass profiles not only of Coma, but of any galaxy cluster.

We can therefore conclude that dwarf galaxies in Coma have radially
anisotropic orbits even close to the cluster center.  This is at
variance with any other type of galaxies (Biviano and Katgert 2004). 
Late-type galaxies do move along radially elongated
orbits but far from the center. It is tempting to interpret
these results in evolutionary terms. 

Dwarf galaxies could be the
remnants of those galaxies that fall into Coma with radial
orbits. Their radial orbits drove them very near the cluster center
where they were morphologically transformed by some physical mechanism
that is effective only near the center (e.g. tidal effects related to
the cluster potential). Galaxies that we still see today as giant
spirals would then be those that did not pass very near the Coma
center and so managed to survive. Hence their orbits cannot be radially strongly
elongated in the central regions (or their pericenters would
be small).

Unfortunately, these results depend on the solution for $N(R_n)$. The
density profile is not well determined because it is only well known
near the cluster center. We have tried to assess the systematics by
forcing the slope of the core model to the value $a=-1$, which
corresponds to the traditional King model (King 1962). The fit is still acceptable,
with a larger core radius than before, $R_c/r_{200}=1.7$ (see
Fig.~\ref{chi2}). Using this new $N(R_n)$ and the same $M(r)$ as before,
we obtain a best-fit constant anisotropy solution of the Jeans equation
$\beta'=1.3 \pm 0.2$ (see Figs.~\ref{f-NR} and \ref{chi2}). This
solution is consistent with the one previously found, but is also
(marginally) consistent with isotropy.

With the current data-set, we have marginally significant evidence that
the dwarf galaxies follow radially anisotropic orbits near the center of
Coma.  Extending the photometric data-set to larger radii would be very
useful to better constrain the number density profile slope, which seems
to play a critical r\^ole in the solution of the Jeans equation.

Moreover, if we arbitrarily split the dwarf galaxy sample in two parts
(below and above R=22.3) a puzzling behaviour appears: dwarf galaxies
with R$\leq$22.3 have a mean velocity of 6721$\pm$350 km/s, while dwarf
galaxies with R$\geq$22.3 have a mean velocity of 7846$\pm$272 km/s. The
two values appear quite different and a Kolmogorov-Smirnov test leads to
a 93$\%$ probability that the two velocity distributions (below and
above R=22.3) are different. The fainter sample has a mean velocity
equivalent to that of the G4 group of Adami et al. (2005a: not sampled
by the VIMOS data). This group is related to the giant galaxy NGC~4911,
so we could expect to have the R$\geq$22.3 dwarf galaxies spatially
correlated with the position of NGC~4911. However, this is not the case:
a Kolmogorov-Smirnov test does not show any evidence for a different
spatial distribution between the two samples of galaxies. An explanation
would be that the NGC~4911 group is losing its faint galaxy population
along its trajectory, spreading this population all over the Coma
cluster. 

The velocity dispersions of the two samples are less different: galaxies
with R$\geq$22.3 have 1516$_{-179}^{+203}$ km/s and galaxies with
R$\leq$22.3 have 1942$_{-230}^{+260}$ km/s, higher than the value
computed in Adami et al. (1998) of 1200 km/s for R$\sim$17.5. This is
not surprising as we deal here with very low mass objects, for which
dynamical relaxation mechanisms are very inefficient in removing initial energy and
then in reducing the velocity dispersion.

\section{Galaxy spectral characteristics}

\begin{table}
\caption{Spectral characteristics of the different stacks. Class is described
  in the text. N is the number of individual
galaxies in each class providing a clean spectrum. log(age yr) is the
logarithm of the stellar population age expressed in years. $\log(M_{\star}/M_{\odot})$
is the logarithm of the stellar mass in solar unit.}
\begin{center}
\begin{tabular}{ccll}
\hline
Class   &  N & log(age yr) & $\log(M_{\star}/M_{\odot})$  \\
\hline
Coma blue        & 15   &     9.13$_{+0.60}^{-0.31}$   &   6.21$_{+0.19}^{-0.14}$           \\ 
Coma red         & 11   &     9.78$_{+0.23}^{-0.30}$   &   6.93$_{+0.20}^{-0.21}$	               \\ 
Coma RS          & 25   &     9.53 $_{+0.38}^{-0.36}$   &   6.66$_{+0.22}^{-0.20}$                \\ 
Coma non X           & 36   &    9.51$_{+0.39}^{-0.35}$  &   6.64$_{+0.22}^{-0.20}$                   \\ 
Coma X       & 28   &    9.46$_{+0.39}^{-0.32}$   &    6.50$_{+0.21}^{-0.17}$                    \\ 
Coma LSB         &  7  &      9.48$_{+0.41}^{-0.38}$   &   6.62$_{+0.25}^{-0.23}$               \\ 
Coma abs         &  31  &     9.51$_{+0.37}^{-0.33}$     & 6.66$_{+0.20}^{-0.18}$                   \\ 
Coma abs R$>$22 &  15  &   9.41$_{+0.42}^{-0.33}$   & 6.39$_{+0.20}^{-0.17}$	               \\
Coma abs R$\leq$22 &  16  &    9.63$_{+0.32}^{-0.37}$    & 7.07$_{+0.26}^{-0.27}$                   \\ 
Coma em+abs      & 26   &      9.45 $_{+0.40}^{-0.33}$    &  6.45$_{+0.21}^{-0.18}$                 \\ 
Coma em+abs R$>$22 & 18   &   9.47$_{+0.39}^{-0.33}$   &  6.39$_{+0.22}^{-0.19}$                       \\ 
Coma em+abs R$\leq$22 &  8  &    9.46$_{+0.44}^{-0.44}$   &  6.93$_{+0.28}^{-0.29}$                 \\ 
Coma em          & 4   &     9.17$_{+0.60}^{-0.38}$   &  6.48$_{+0.24}^{-0.22}$                     \\ 
Non Coma abs         & 44   &       9.48 $_{+0.40}^{-0.35}$     &  /              \\ 
Non Coma abs R$>$22 &  31  &   9.50 $_{+0.40}^{-0.37}$     &    /                  \\ 
Non Coma abs R$\leq$22 &  13  &    9.29$_{+0.57}^{-0.59}$      &   /                   \\ 
Non Coma em+abs      &  39  &      9.12$_{+0.67}^{-0.36}$     &   /                    \\ 
Non Coma em+abs R$>$22 & 25   &     9.20$_{+0.62}^{-0.41}$	   &  /                          \\ 
Non Coma em+abs R$\leq$22 & 14   &     9.07 $_{+0.73}^{-0.45}$     &   /                  \\ 
Non Coma em          &  4  &     9.05$_{+0.81}^{-0.68}$       &   /                   \\ 
Non Coma LSB         &  4  &     9.76$_{+0.28}^{-0.39}$        &  /                   \\ 
\hline
\end{tabular}
\label{fitspectral}
\end{center}
\end{table}

\subsection{Method}

We have produced stacked spectra by computing a weighted mean of the
individual clean spectra available, using the $R$-band magnitude as a
weight.  We measured spectral indices from these stacked spectra and fit
them, together with the stacked spectral energy distributions in
the $u^*BVRI$ bands, with a library of stellar population models. The
absorption indices which were used in the fit are Lick\_G4300,
Lick\_Mgb, Lick\_Fe5270, Lick\_NaD, BH\_G, and BH\_Mgg. We also used the
4000~\AA{} break: Dn4000.

The fit was performed in a bayesian approach, namely we computed for
each stack the probability distribution function (hereafter PDF) of each
desired parameter, given an input library of 100~000 models with uniform
coverage in their physical parameters (see Salim et al.  2005, Walcher
et al. 2008, Lamareille et al. 2009). These models include in particular
complex star formation histories. The IMF is that of Chabrier (2003).

We thus derived for each stack the age of the oldest stellar populations,
the stellar mass, the dust attenuation, and the stellar metallicity.

\subsection{Properties of Coma cluster dwarf galaxies as a function of color:
  red, blue, and red-sequence galaxies.}

\begin{figure*}
\begin{centering}
\includegraphics[width=0.5\linewidth]{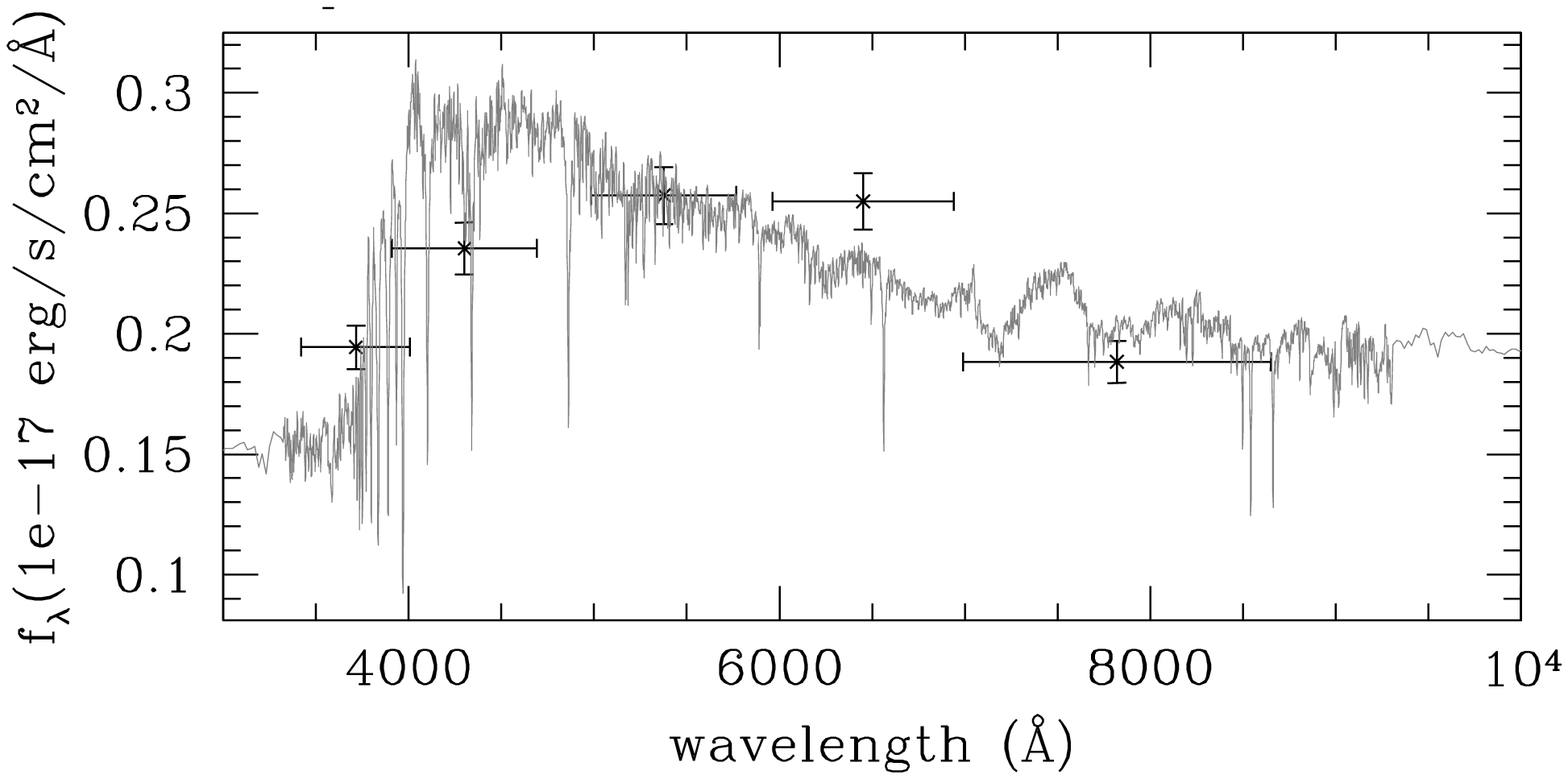}\includegraphics[width=0.25\linewidth]{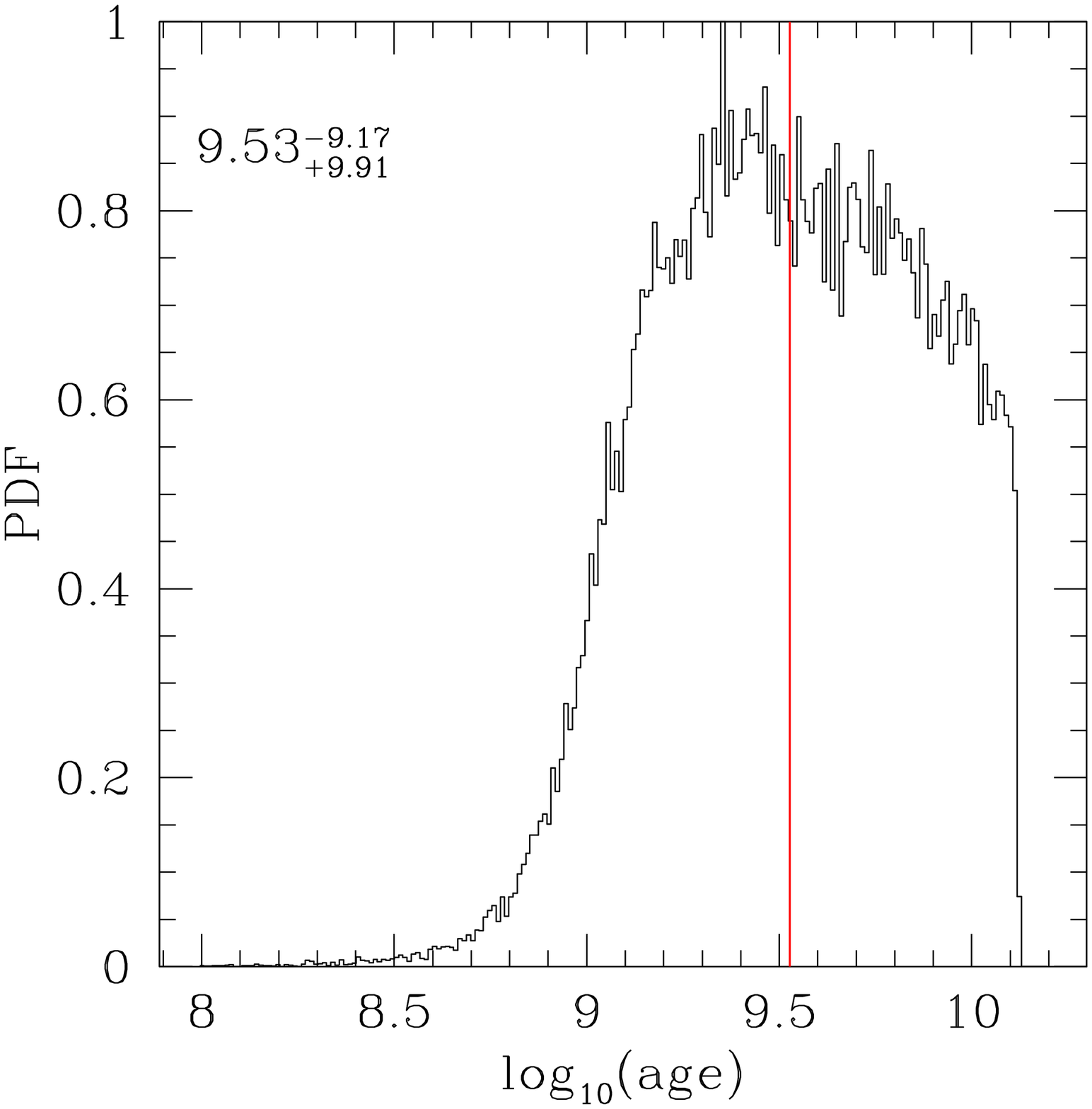}\includegraphics[width=0.25\linewidth]{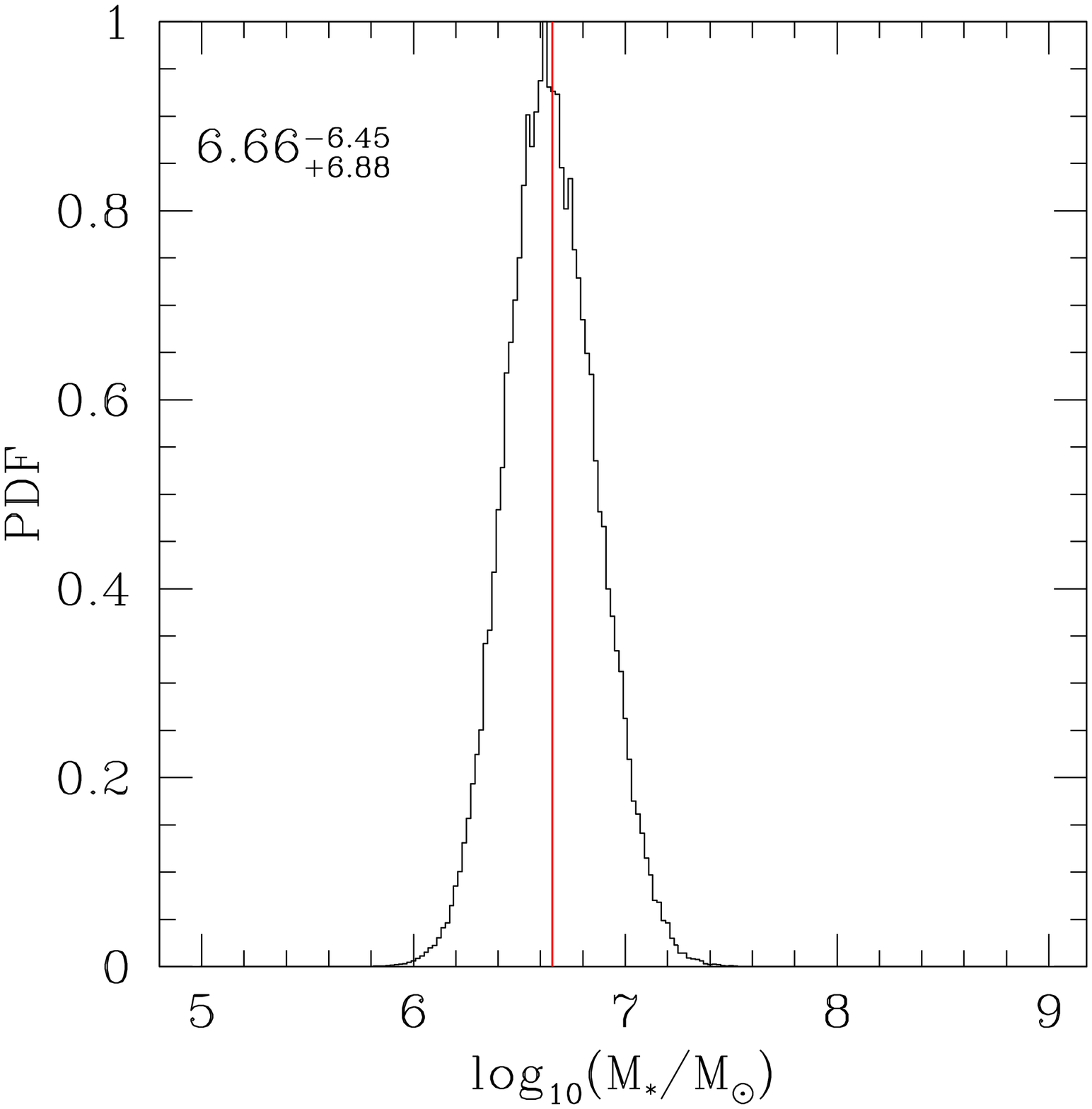}\\
\includegraphics[width=0.5\linewidth]{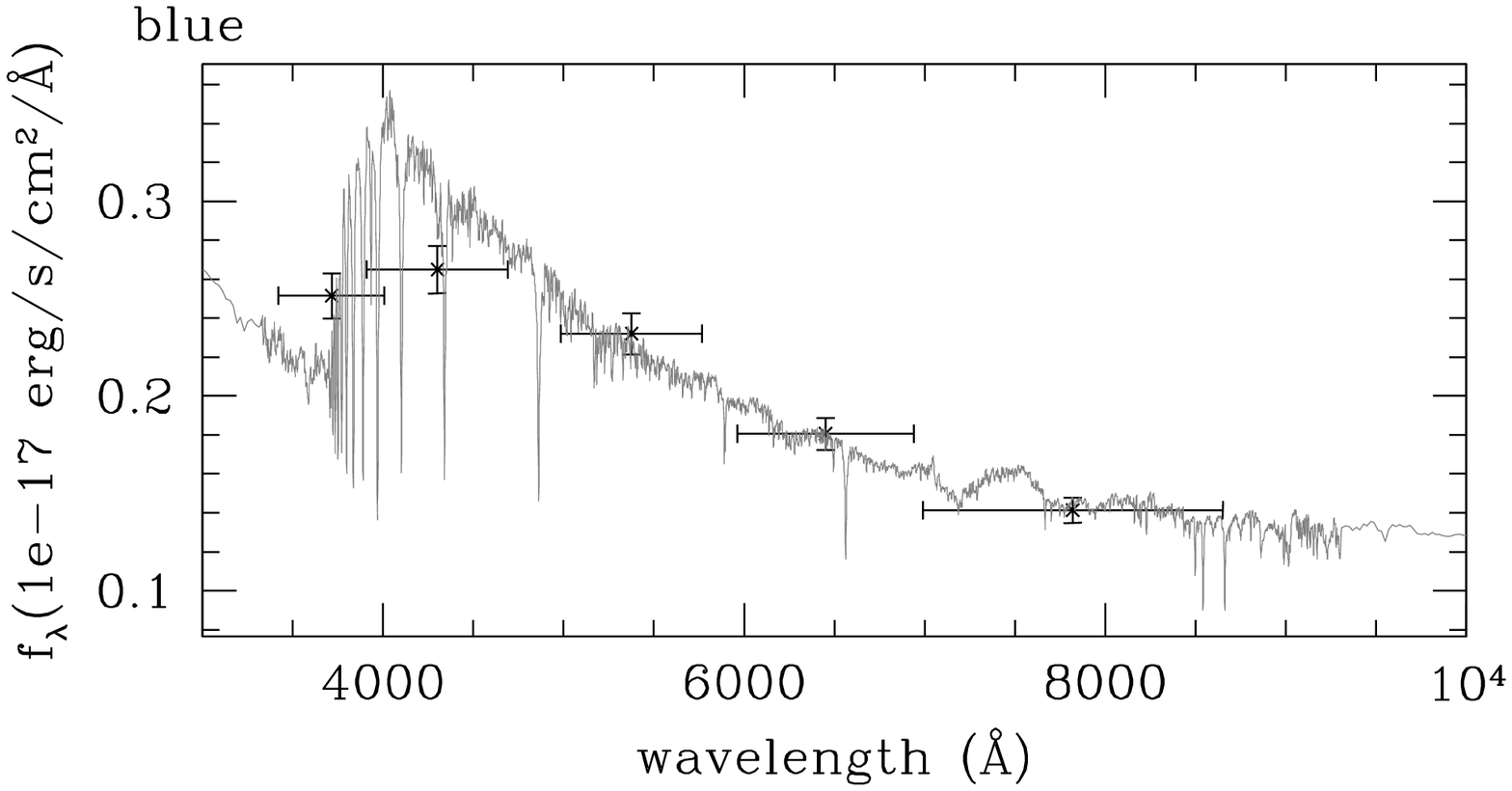}\includegraphics[width=0.25\linewidth]{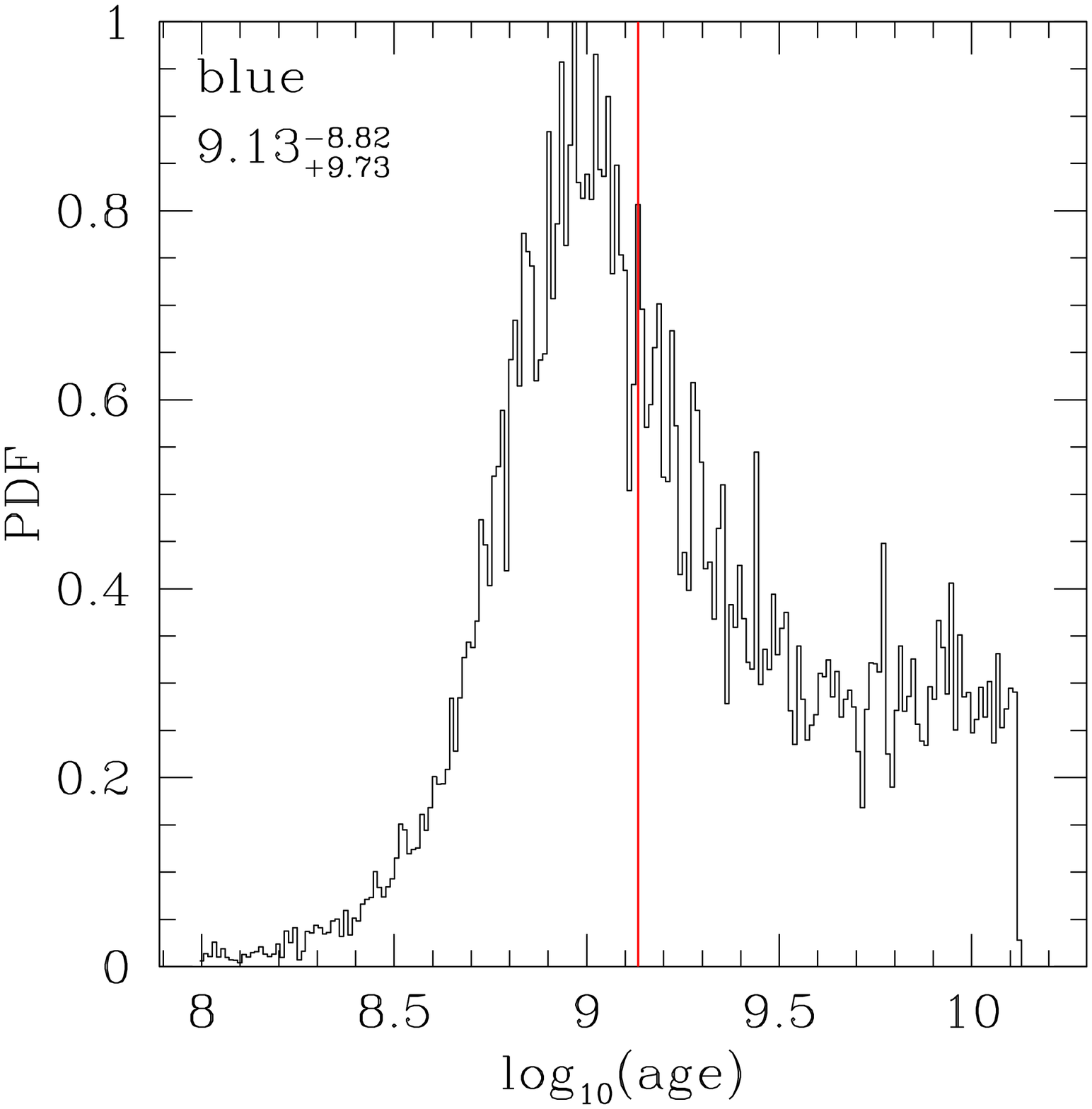}\includegraphics[width=0.25\linewidth]{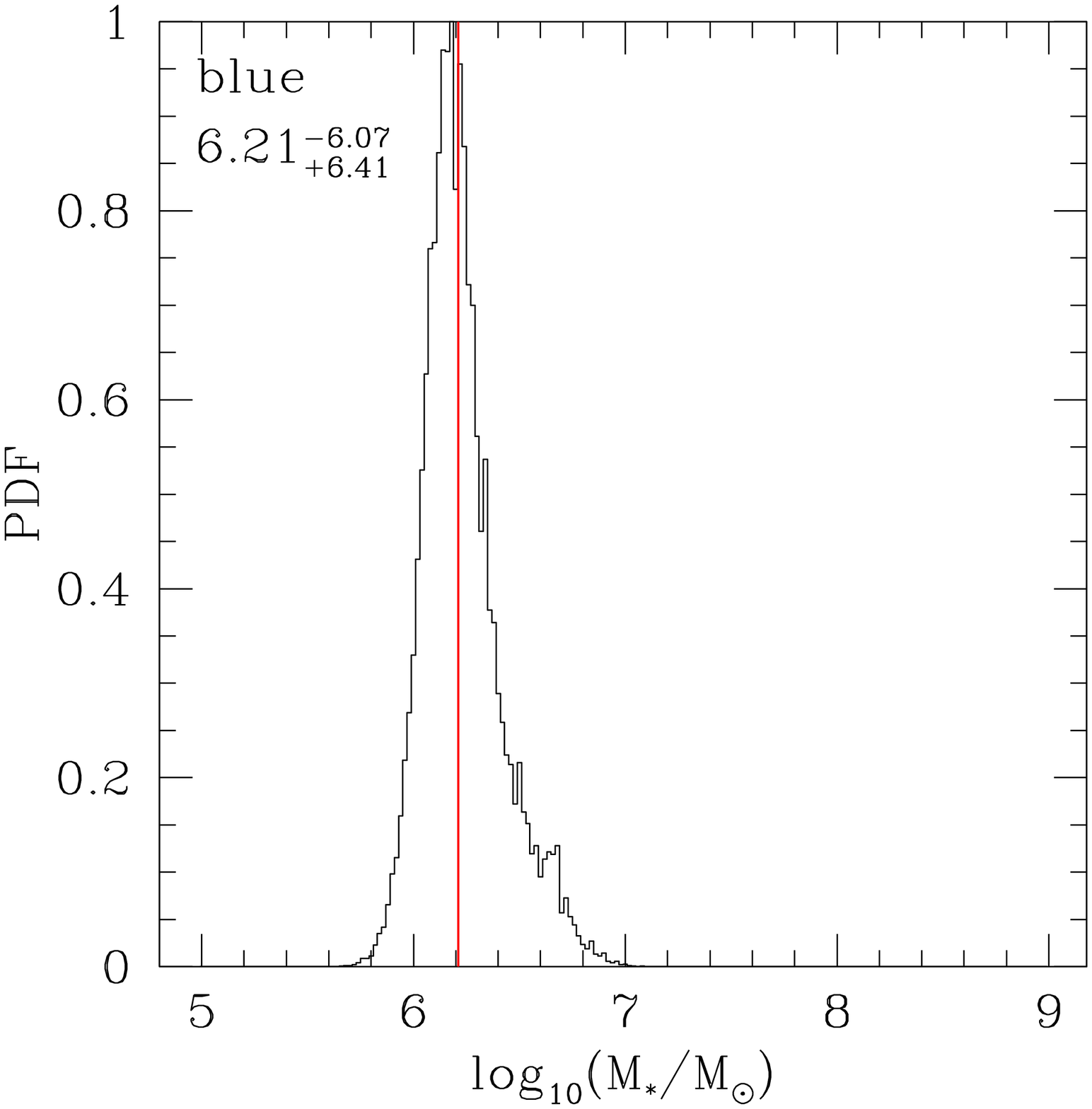}\\
\includegraphics[width=0.5\linewidth]{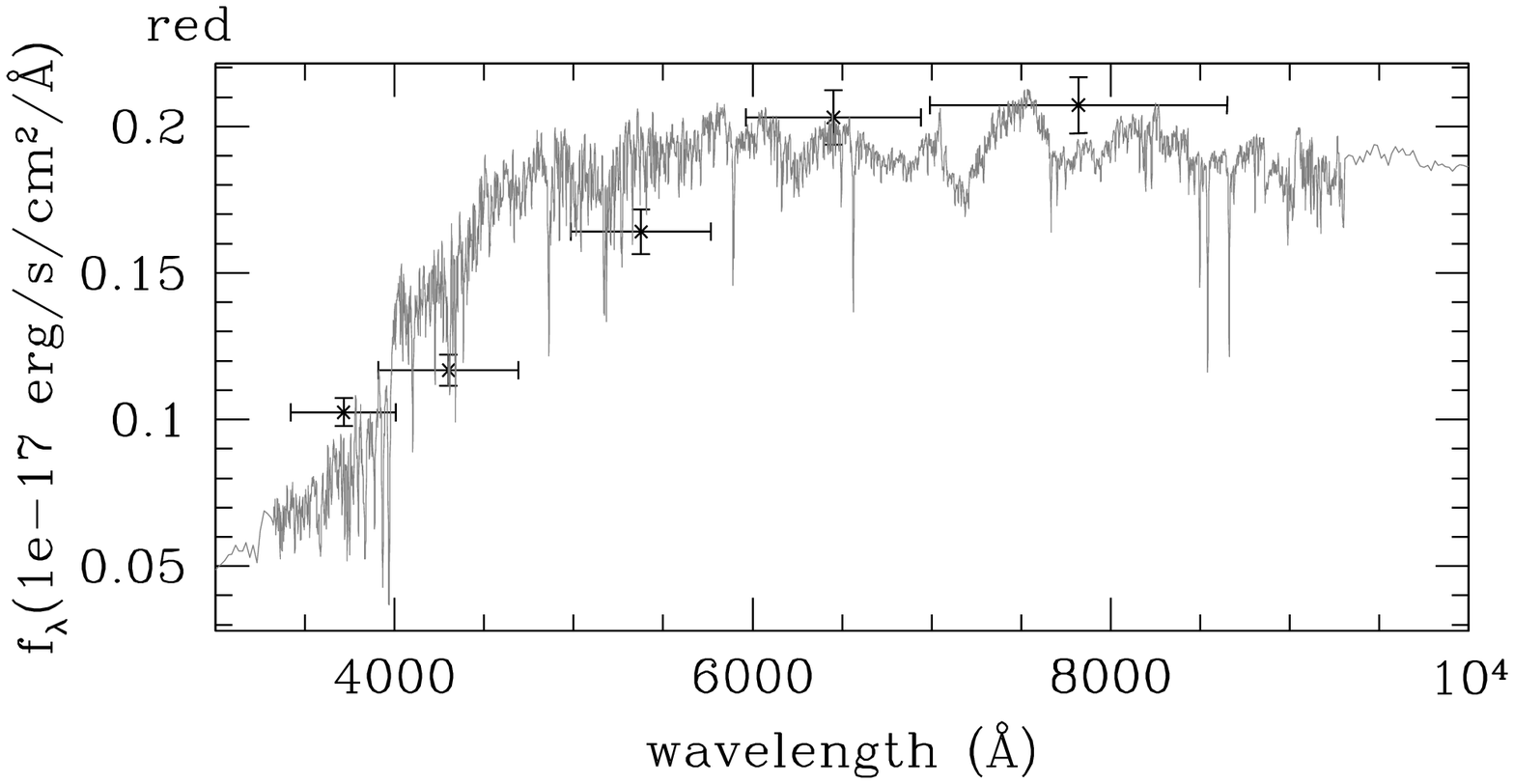}\includegraphics[width=0.25\linewidth]{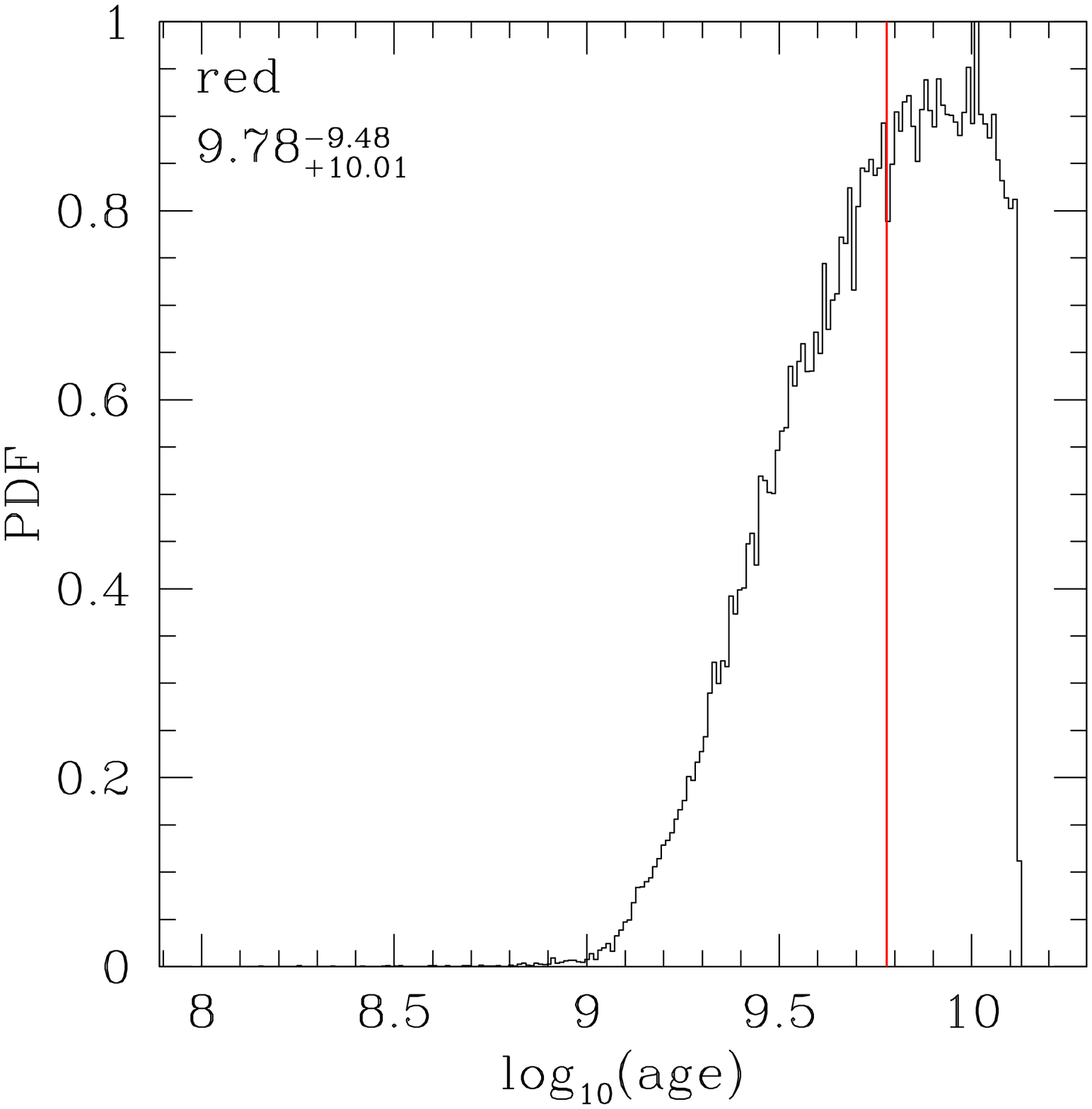}\includegraphics[width=0.25\linewidth]{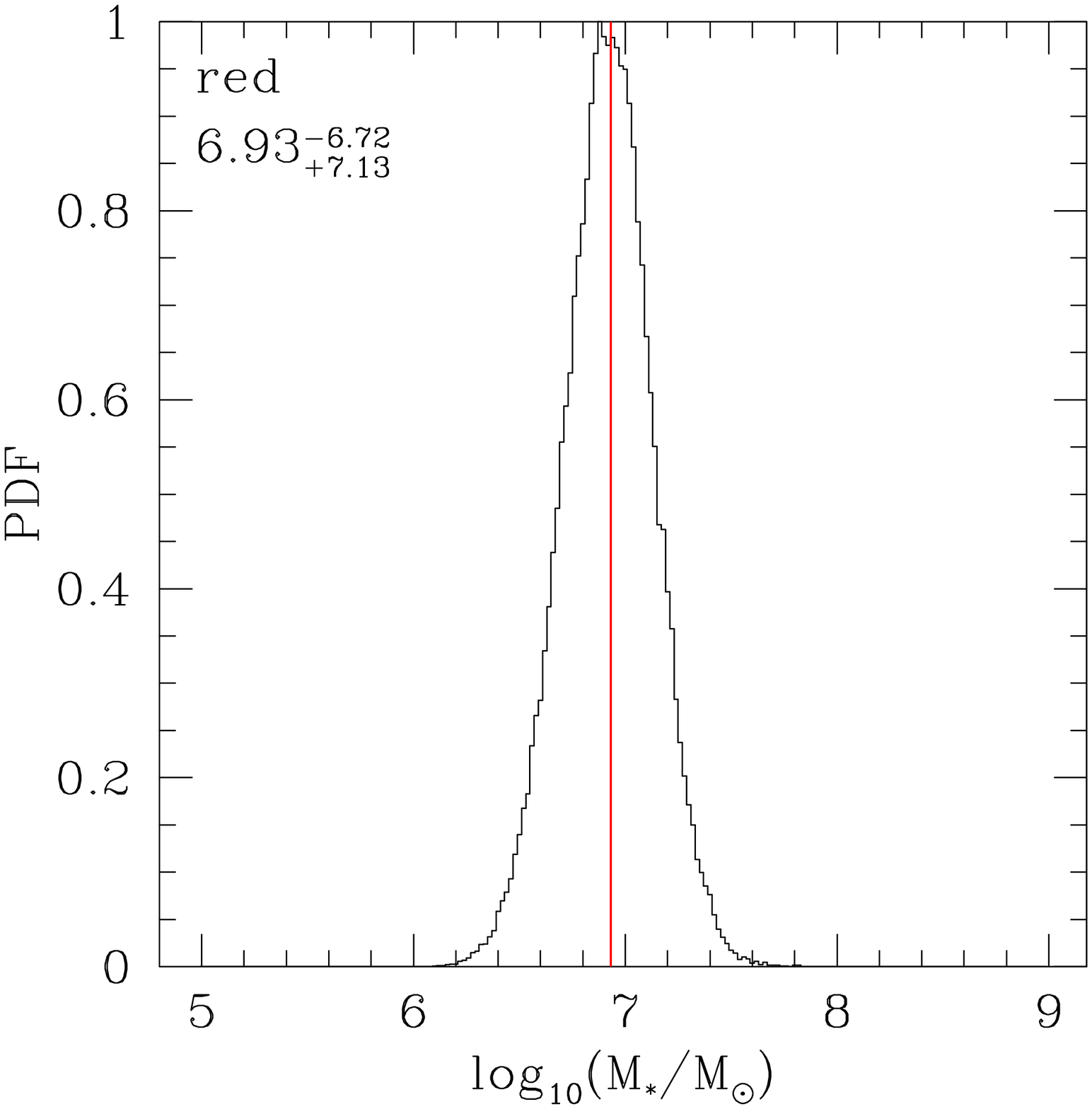}
\par\end{centering} \caption{Spectral energy distributions (left) for
three stacks of Coma cluster dwarf galaxies: red sequence population,
blue galaxies, and red galaxies (top to bottom). The points with error
bars show the $u^*BVRI$ magnitudes.  The overlaid spectrum shows the
mean stellar population model. Probability distribution functions are
shown to the right for the ages of the oldest stellar populations and
for the stellar masses (both in logarithmic scale) for the three same 
stacks. The vertical line shows the median estimate, which is also given
in the plot. Ages are given in years.}  \label{figsed}
\end{figure*}

In this analysis we consider three different classes of dwarf galaxies:
red galaxies, red-sequence galaxies, and blue galaxies (see section
7.2).  Fig.~\ref{figsed} shows the spectral energy distributions and the
mean stellar population model for these classes: red-sequence
population, blue galaxies, and red galaxies. This figure also shows the
PDF for the ages of the oldest stellar populations and the stellar
masses.

From blue to red sequence, to red galaxies respectively, the ages of the
Coma cluster dwarf galaxies range from $\log(\mbox{age})=9.13$ to 9.53
and 9.78.  The blue galaxies therefore seem to have a young stellar population
($\sim$1.3 Gyrs) while red galaxies appear to be older objects ($\sim$6
Gyrs). The stellar masses range from $\log(M_{\star}/M_{\odot})=6.21$ to
6.66 and 6.93. Red galaxies are therefore more massive (regarding their
stellar mass) than blue galaxies. We find here for faint dwarf galaxies
the same well known tendencies observed for giant galaxies.

Dust attenuation and stellar metallicity do not vary significantly 
and are not given in Table~\ref{fitspectral}. Metallicities are
compatible with solar values, in good agreement with the assumption of
Adami et al. (2009b).

\subsection{Comparison with non-Coma cluster dwarf galaxies}

\begin{figure*}
\begin{centering}
\includegraphics[width=0.25\linewidth]{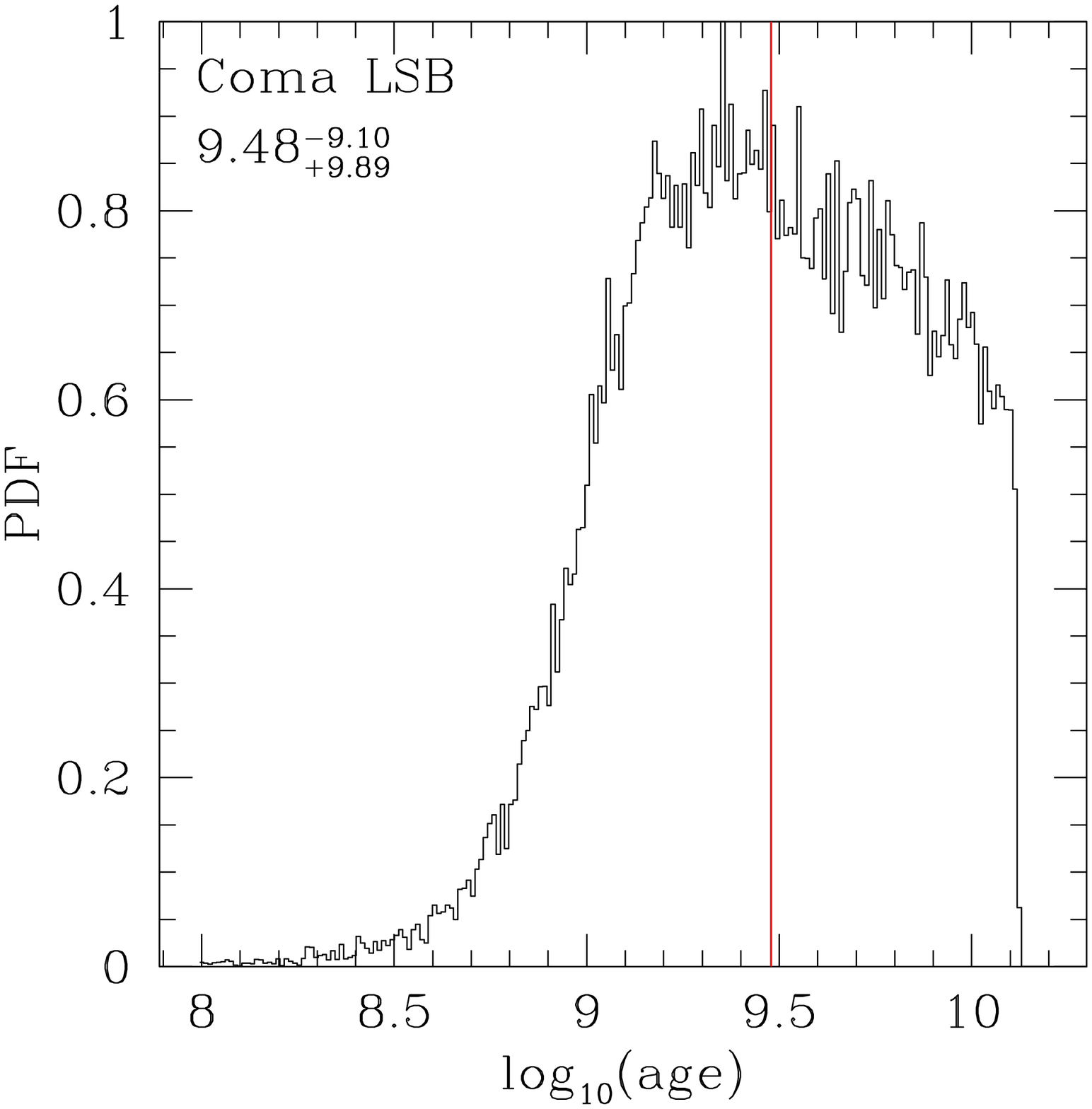}\includegraphics[width=0.25\linewidth]{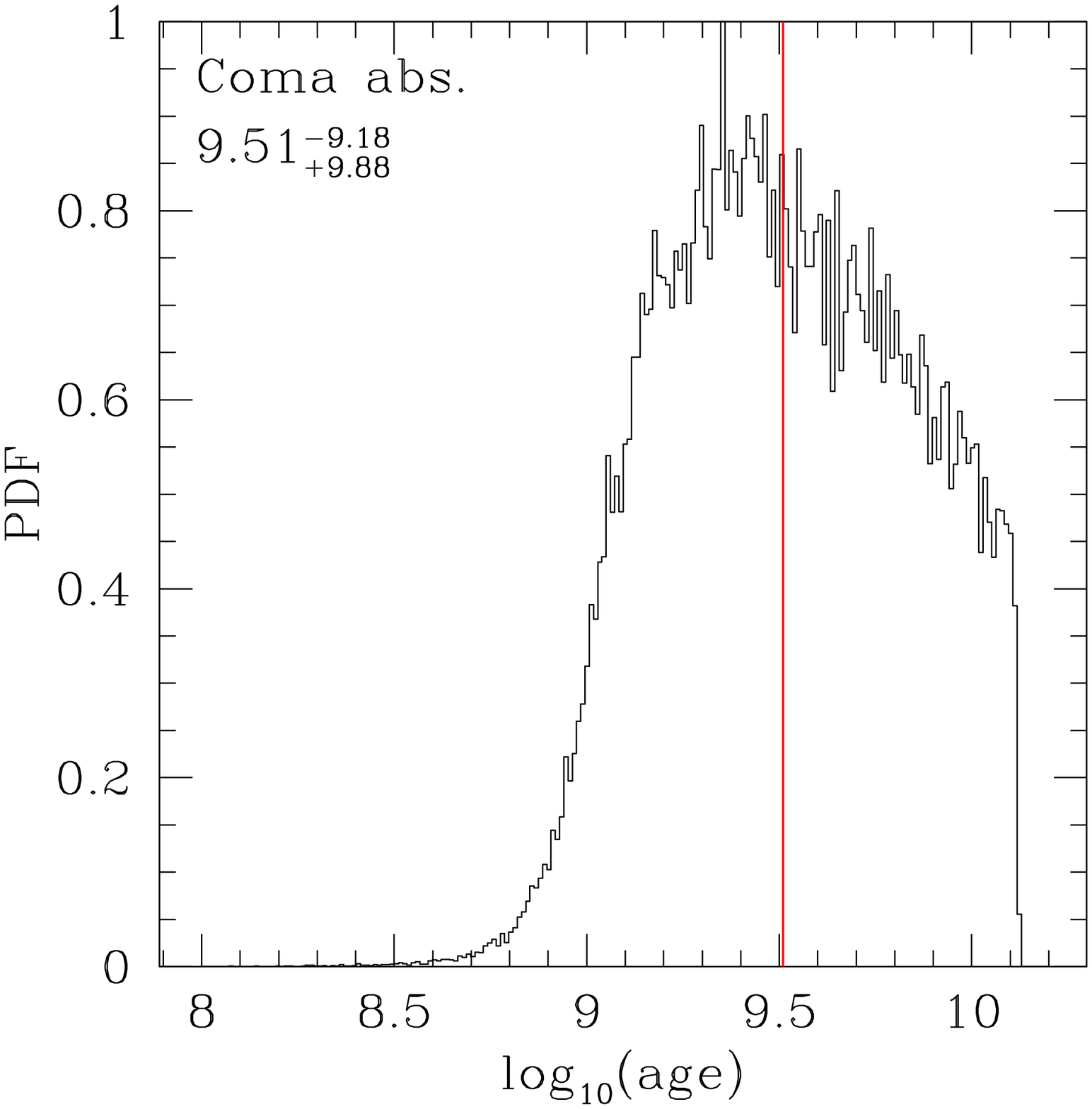}\includegraphics[width=0.25\linewidth]{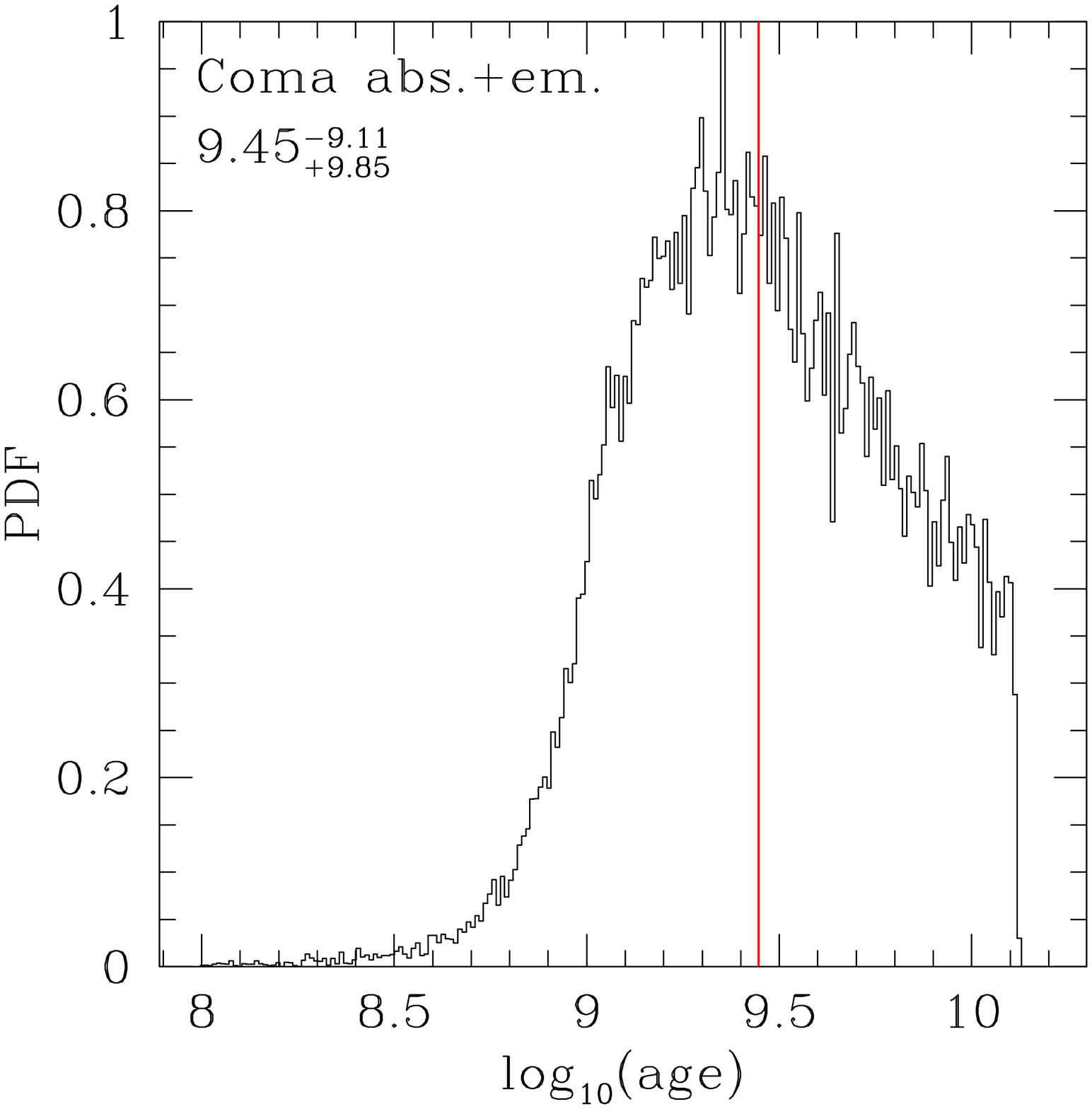}\includegraphics[width=0.25\linewidth]{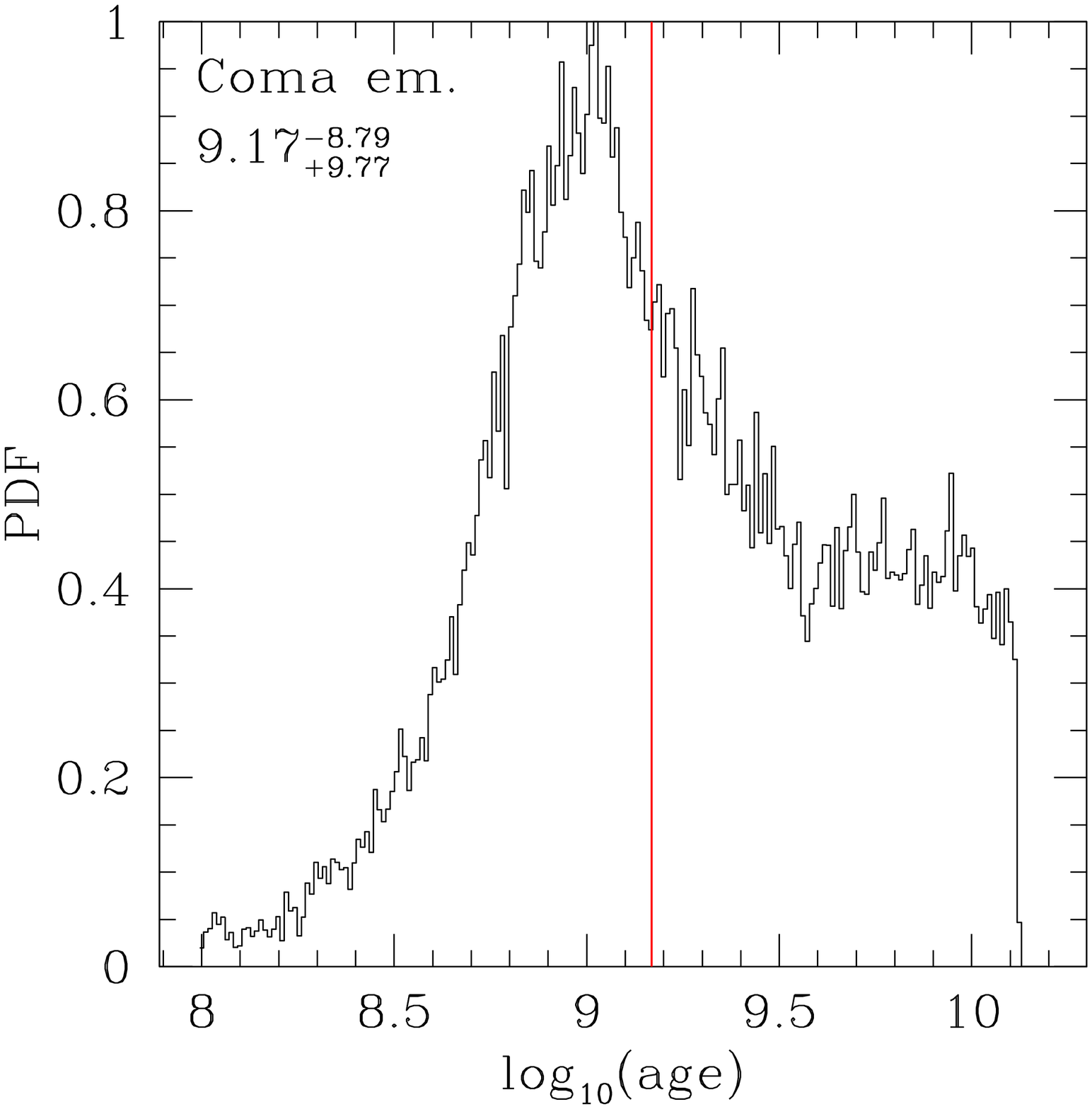}\\
\includegraphics[width=0.25\linewidth]{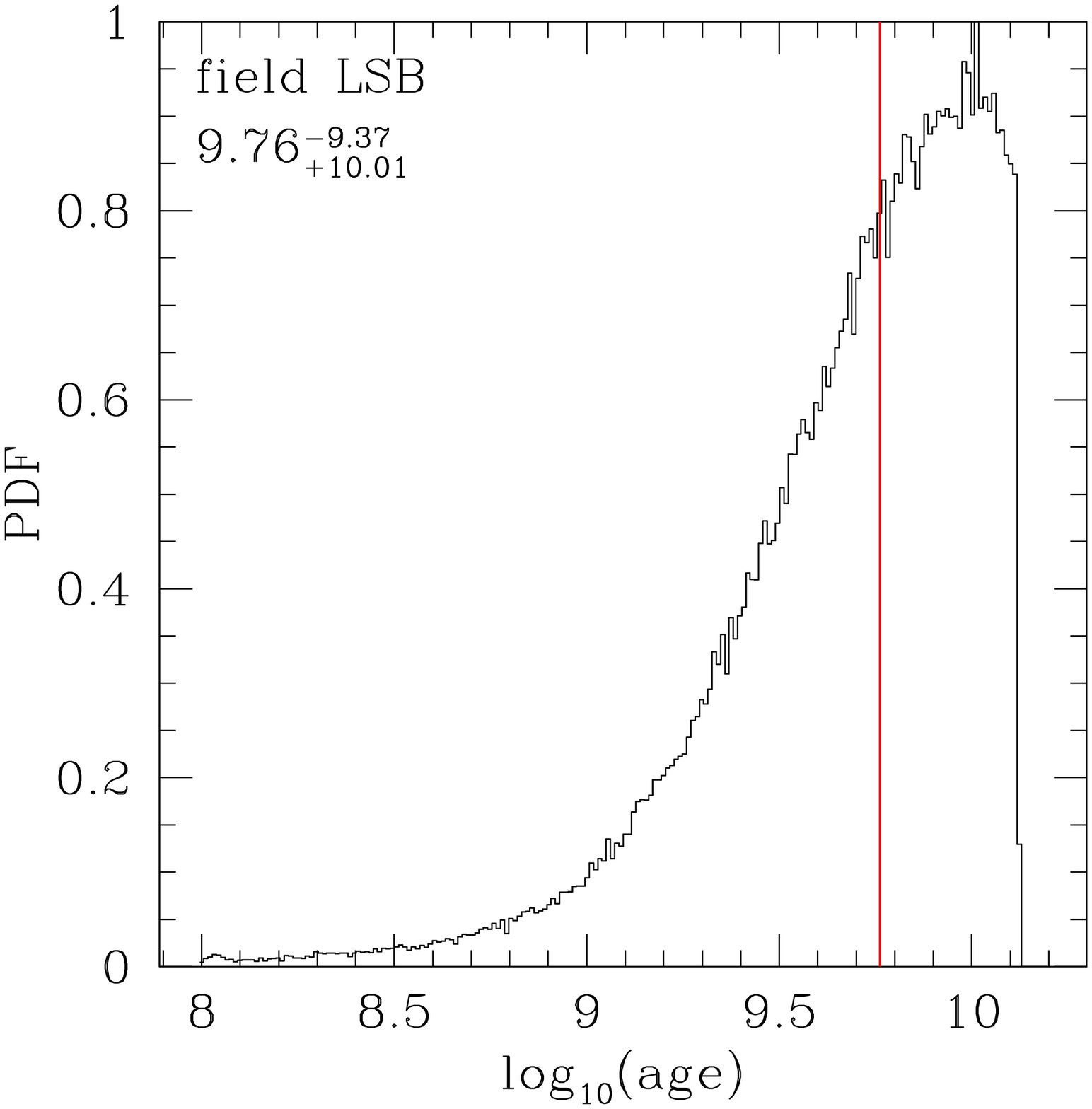}\includegraphics[width=0.25\linewidth]{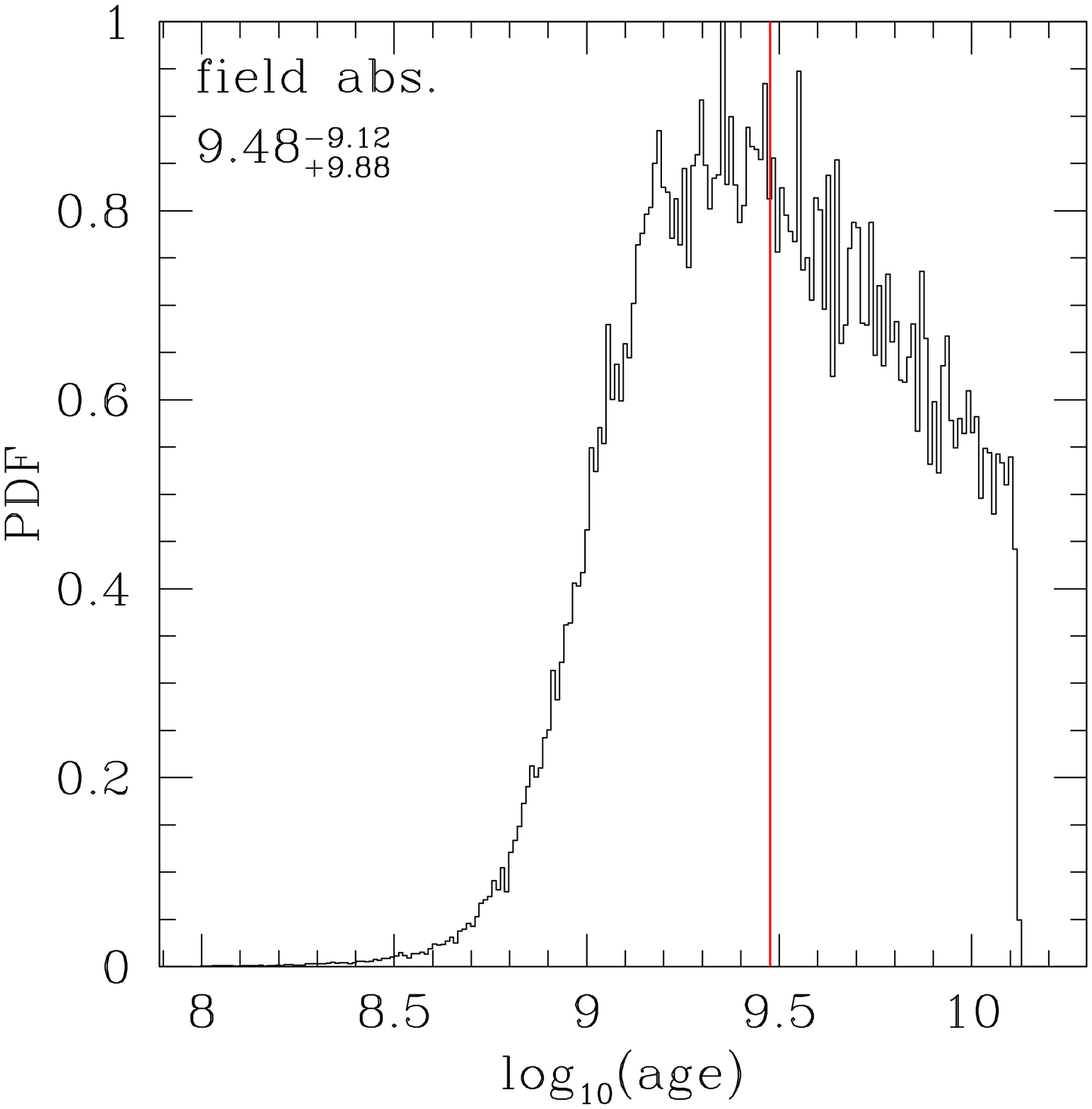}\includegraphics[width=0.25\linewidth]{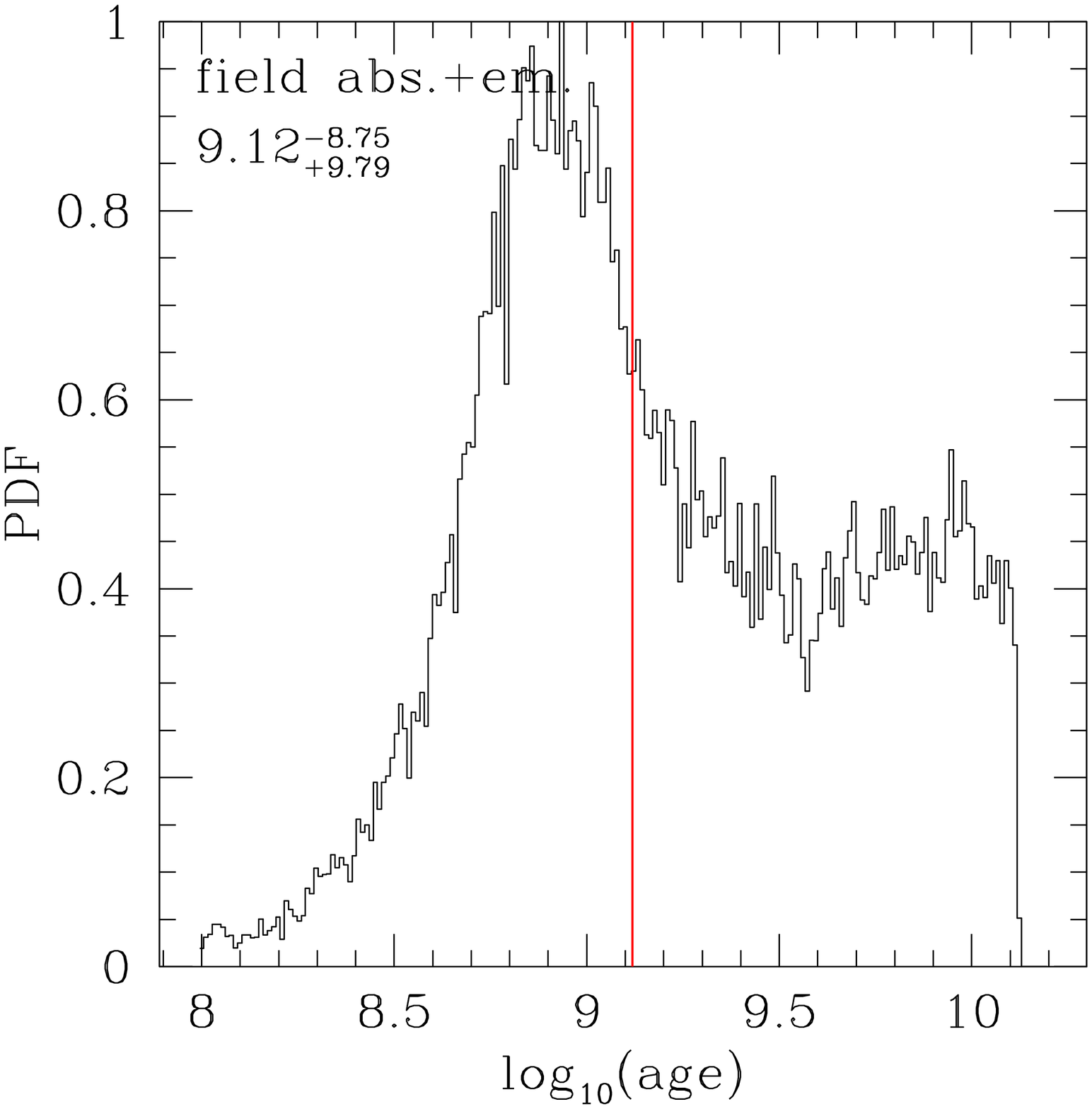}\includegraphics[width=0.25\linewidth]{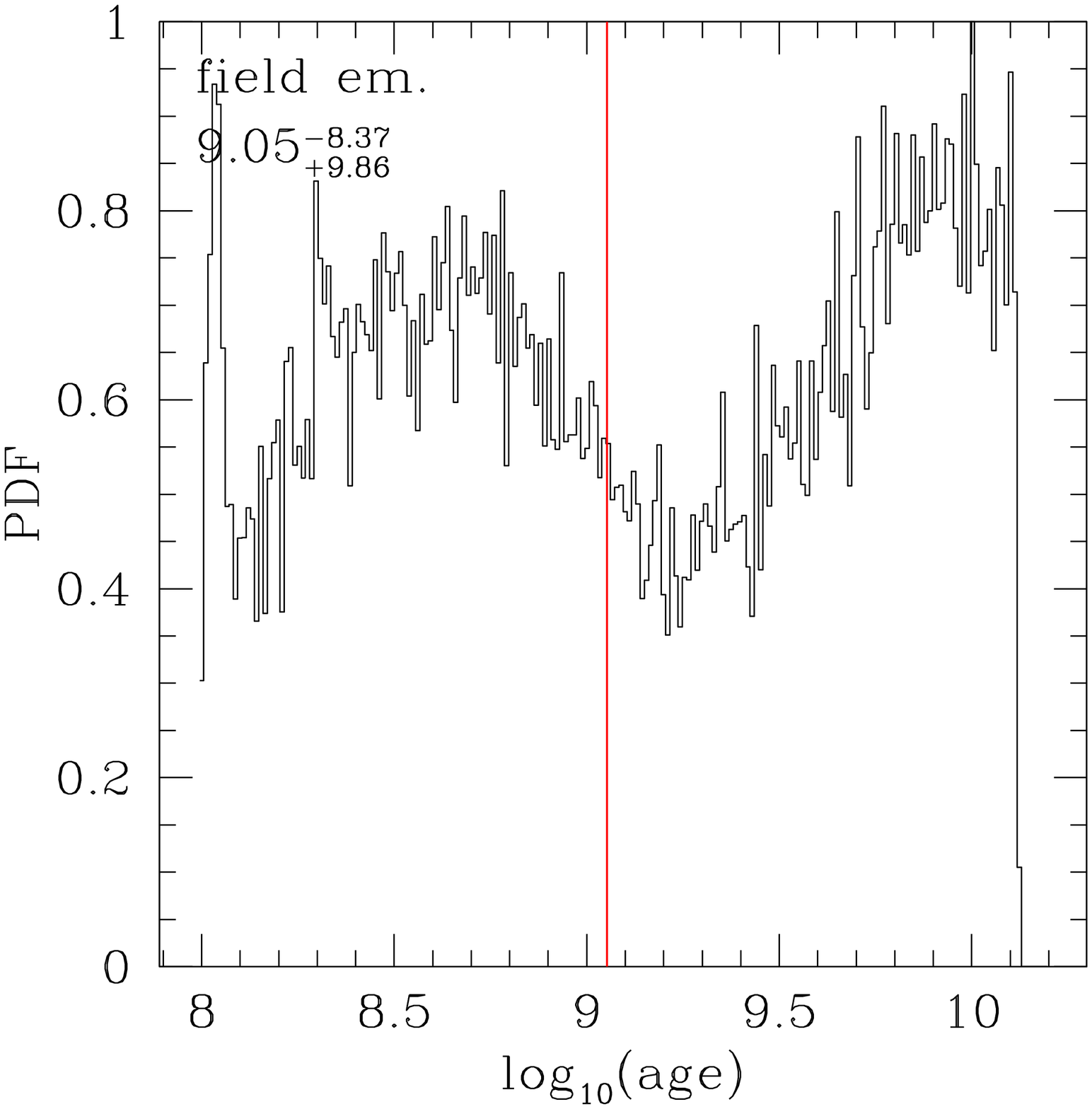}
\par\end{centering}

\caption{Probability distribution functions for the ages of the oldest
stellar populations (both in logarithmic scales) for dwarf galaxies
inside (top) or outside (bottom) the Coma cluster. From left to right:
low surface brightness, absorption-line, absorption- plus emission-line,
and emission-line galaxies. The vertical line shows the median estimate,
which is also written in each plot. Ages are given in years.}

\label{figpdf}
\end{figure*}

In this analysis, we classify our dwarf galaxies in four different
classes: low surface brightness galaxies, absorption-line galaxies,
absorption- plus emission-line galaxies, emission-line galaxies. These
correspond to the {\it Coma LSB}, {\it Coma abs}, {\it Coma em+abs}, and
{\it Coma em} lines in Table~\ref{fitspectral} for the Coma cluster and
to {\it Non Coma LSB}, {\it Non Coma abs}, {\it Non Coma em+abs}, and
{\it Non Coma em} lines in Table~\ref{fitspectral} for the z$\leq$0.1
non-Coma cluster galaxies. These galaxies act as a comparison sample,
not being subject to the strong influence of the Coma cluster and being
at sufficiently low redshift not to exhibit strong evolutionary
effects. Fig.~\ref{figpdf} shows the PDF for the ages of the oldest
stellar populations for these four classes (from left to right) and for
galaxies inside and outside the Coma cluster (top to bottom). We observe
two trends. First, dwarf galaxies show decreasing mean ages from
absorption-line to emission-line galaxies.  We note that this trend is
also associated with decreasing stellar masses. Second, the Coma cluster
dwarf galaxies seem on average older than field dwarf galaxies, even if
this trend is not significant. Conversely, low surface brightness
galaxies seem to follow the opposite trend: they are
barely-significantly older outside the Coma cluster than inside it,
where they have ages similar to absorption-line galaxies. There is no
significant evidence of an age trend, although, taken at face value, the mean
ages would indicate a different origin for at
least part of the LSBs in the field and in the Coma cluster. This would
be in good agreement with an interaction-induced origin for part of the
Coma LSBs (see e.g. Adami et al. 2009b).

Finally, it is interesting to note that the metallicity of the Coma
intracluster medium is 0.2$Z_{\odot}$ (Strigari et al. 2008), clearly
less metal--rich than the dwarf galaxies considered here.

\subsection{Influence of the intracluster medium: X-ray substructures.}

We check in this section if the intracluster medium X-ray substructures have
an influence on the faint Coma dwarf galaxies (R$\geq$21). Galaxies inside
X-ray substructures of Neumann et al. (2003: $Coma~X$ line in
Table~\ref{fitspectral}) and outside of these substructures ($Coma~non~X$ line in
Table~\ref{fitspectral}) have very similar properties without any significant differences given our
uncertainties. This suggests that whichever influence the cluster substructures may have on the
faint dwarf galaxies, this does not occur via the hot gas attached to these
substructures.

\subsection{Magnitude versus spectral properties}

We investigate in this section the possible relation between galaxy
magnitudes and their spectral properties within the main magnitude range covered
by 
the VIMOS data (R=[22,23]). 

Splitting our Coma sample into R$\leq$22 (the {\it Coma abs R$\leq$22}, and
{\it Coma em+abs R$\leq$22} lines in Table~\ref{fitspectral}) and R$>$22
(the {\it Coma abs R$>$22}, and {\it Coma em+abs R$>$22} lines in
Table~\ref{fitspectral}), we show that the ages of galaxies of a given
class (pure emission, emission-absorption, or pure absorption spectral
features) are independent of their luminosities.  Faint and bright
galaxies have very similar ages for a given class (pure emission,
emission-absorption, or pure absorption spectral features).

Not surprisingly, pure absorption line R$\geq$22 galaxies have smaller
stellar masses than pure absorption line R$\leq$22 galaxies . We have
the same tendency for R$\geq$22 and R$\leq$22 emission-absorption
galaxies.  There is no clear correlation between the stellar mass
(expressed in terms of magnitude) and the age.

\begin{figure}
\centering \mbox{\psfig{figure=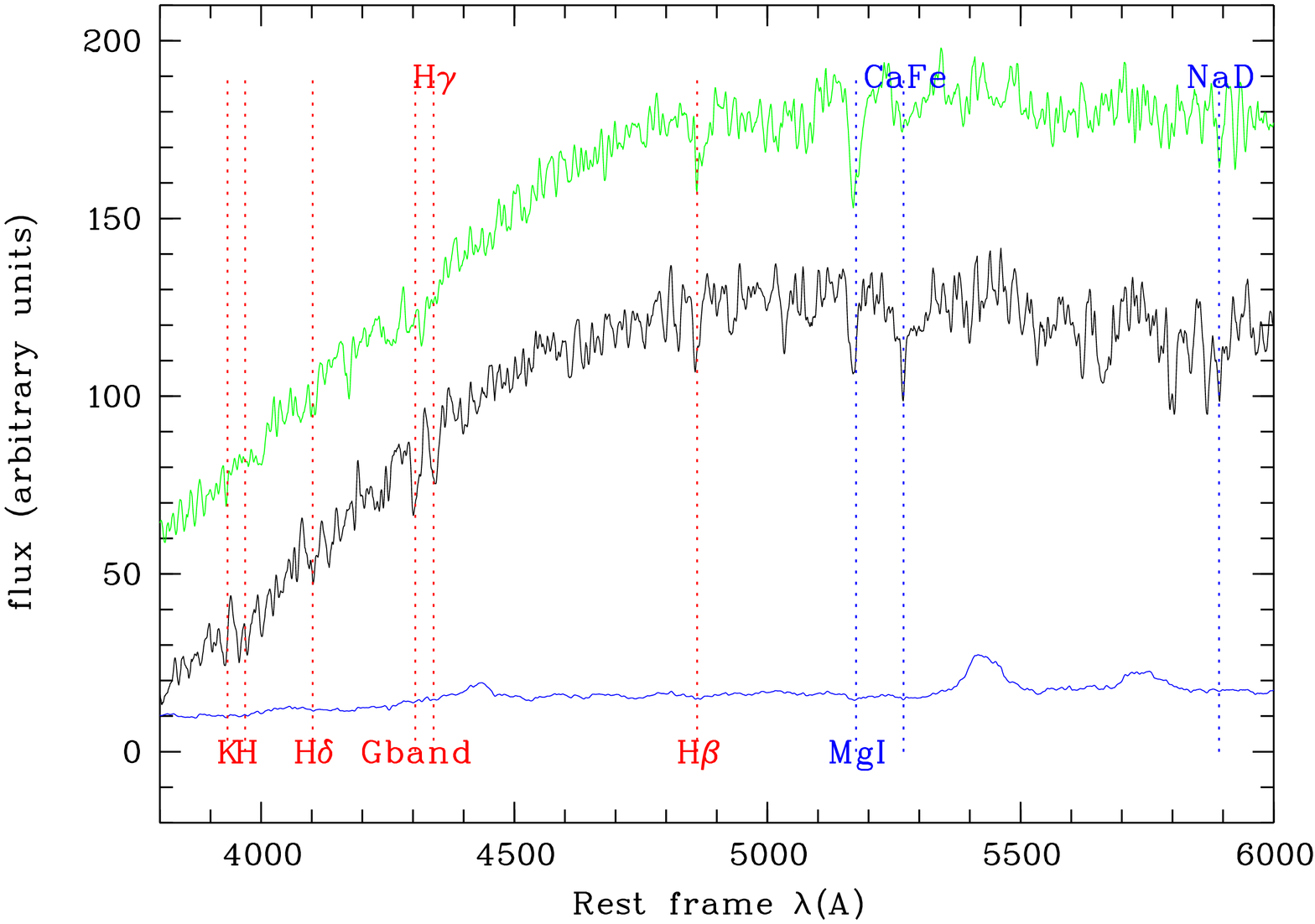,width=8cm,angle=270}}
\centering \mbox{\psfig{figure=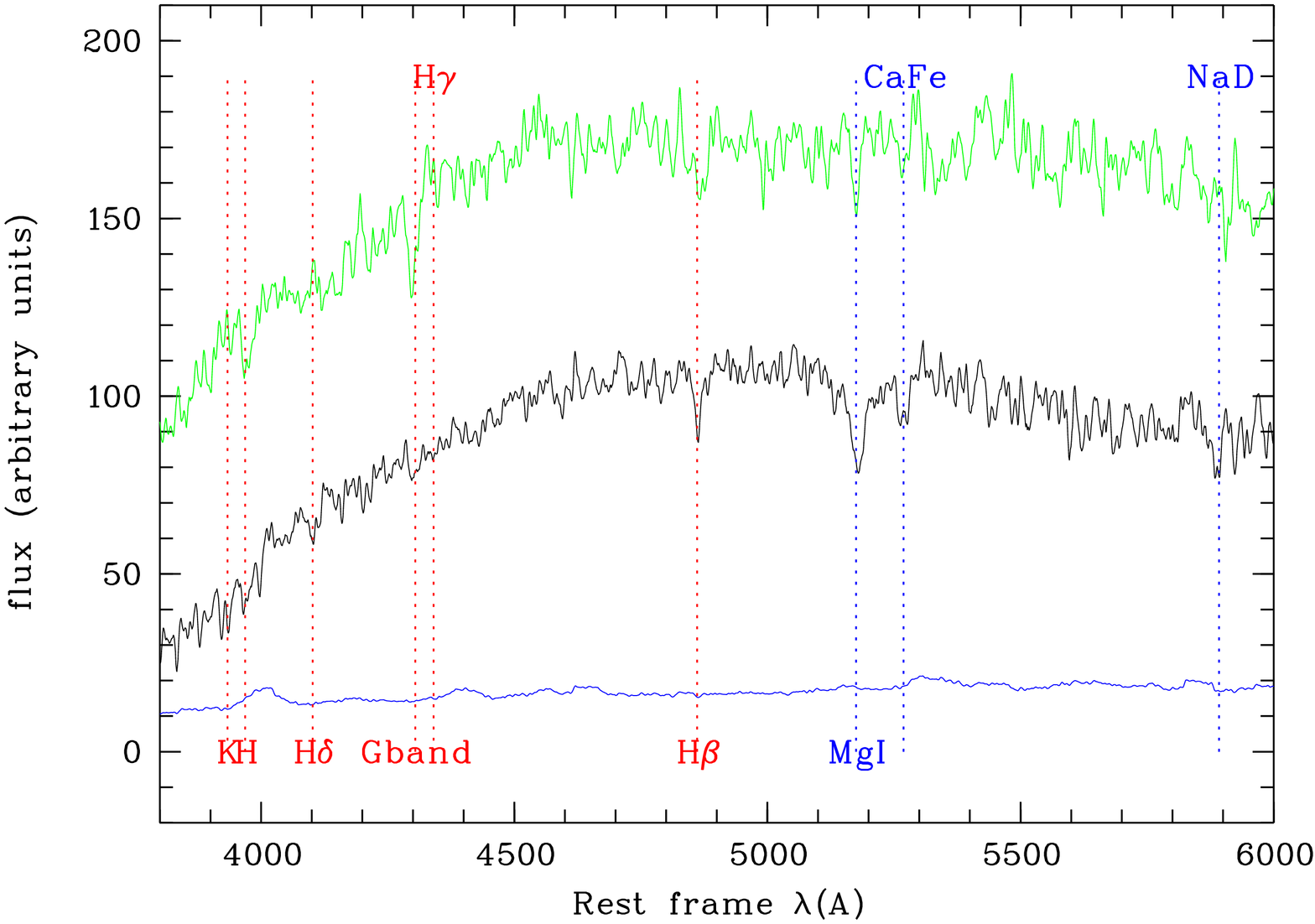,width=8cm,angle=270}}
\caption[]{Upper figure: Coma cluster: composite spectra of the
R$\leq$22 Coma pure-absorption line galaxies (upper spectrum in green),
of the R$\geq$22 Coma pure-absorption line galaxies (lower spectrum in
black), and of the noise level of all the pure-absorption line galaxies
(in blue).  Lower figure: same for z$\leq$0.1 non-Coma galaxies.}
\label{fig:spectr}
\end{figure}

A more detailed analysis (Fig.~\ref{fig:spectr}) however shows
some differences between Coma pure absorption line galaxies at
R$\leq$22 and R$\geq$22. R$\geq$22 galaxies exhibit among others H$\&$K,
and G-band lines, while such lines are only barely visible in R$\leq$22
galaxies.  Non-Coma z$\leq$0.1 galaxies do not show the same trend: they
exhibit similarly strong absorption lines whatever the magnitude. 
An explanation would be that the faintest galaxies (which are qualitatively
similar to Non-Coma z$\leq$0.1 galaxies) were recently injected inside
the Coma cluster directly from the field, perhaps following the NGC~4911
infalling group (see section 5.2).

\section{Ultra Compact Dwarf candidates}

In the previous section, we discussed several Coma cluster galaxy
subclasses, including faint low surface brightness galaxies. We now
investigate the possibility to detect other extreme surface brightness
objects, the so-called Ultra Compact Dwarf galaxies (UCD hereafter).
These objects are characterized by both a very small diameter and a high
surface brightness, being quite similar to stars in terms of luminosity
profile. Only a small number of these galaxies are presently known (see
e.g. Phillipps et al. 2001 or more recent works by Chilingarian et
al. 2008 or Price et al. 2009). For example Price et al. (2009) found
only 3 UCDs in the Coma cluster from an HST-ACS and spectroscopic
analysis (their sample basically cover the B=18-22 magnitude range). Any
additional candidate can therefore significantly contribute to the
study of this class of extreme objects.

We adopted a similar search strategy as in Price et al. (2009) but with a
much fainter magnitude search range: our spectroscopic sample basically starts
where the Price et al. (2009) sample ends. Our photometric data are ground based observations,
so we will be unable to finalize a UCD sample, but we can provide some
candidates which will be offered to the community for validation via
space-based observations.

\subsection{Imaging selection}

\begin{figure}
\centering \mbox{\psfig{figure=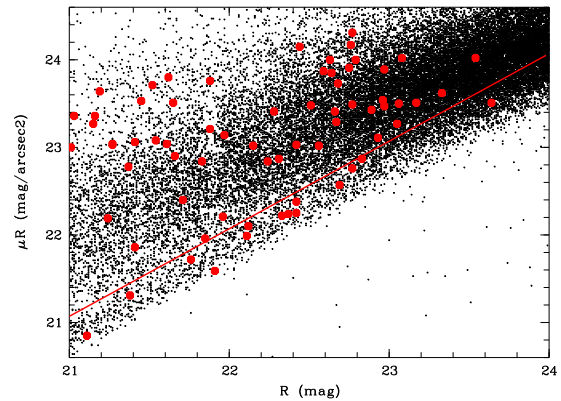,width=8cm,angle=270}}
\caption[]{Central surface brighness versus magnitude diagram for
galaxies. Points are all the galaxies detected along the Coma cluster
line of sight in our survey. Red filled circles are spectroscopically
confirmed cluster members (reliability flag $\geq$ 2). The red inclined
line is the adopted limit for the star sequence. Red filled circles
below this line are a priori UCD candidates (see text).}
\label{fig:UCD1}
\end{figure}

UCDs are a priori located in the star sequence in a surface brightness
versus magnitude diagram. We therefore selected all galaxies
spectroscopically confirmed as Coma cluster members and located in the
star sequence as shown in Fig.~\ref{fig:UCD1}.

\begin{figure}
\centering \mbox{\psfig{figure=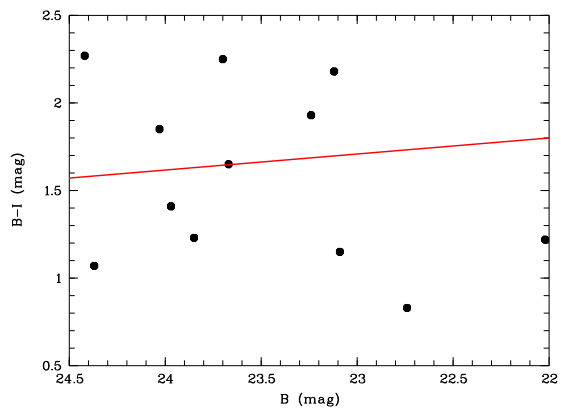,width=8cm,angle=270}}
\caption[]{B-I color versus B magnitude for the galaxies below the star
line in Fig.~\ref{fig:UCD1}. The red inclined line is the lower limit
for a galaxy to be a UCD in the Price et al. (2009) Fig. 2.}
\label{fig:UCD2}
\end{figure}

Next, we used a color criterium in order to refine the list of UCD
candidates. Similarly to Price et al. (2009), we placed in a B-I versus
B diagram the galaxies below the star line in
Fig.~\ref{fig:UCD1}. Fig.~\ref{fig:UCD2} then allows us to reduce our
UCD candidate list to five objects at B-I$\geq$1.8 (see Price et
al. (2009) for details). These five objects are therefore compact enough
and red enough to be potential UCD galaxies.

\subsection{Spectroscopy}

\begin{figure}
\centering \mbox{\psfig{figure=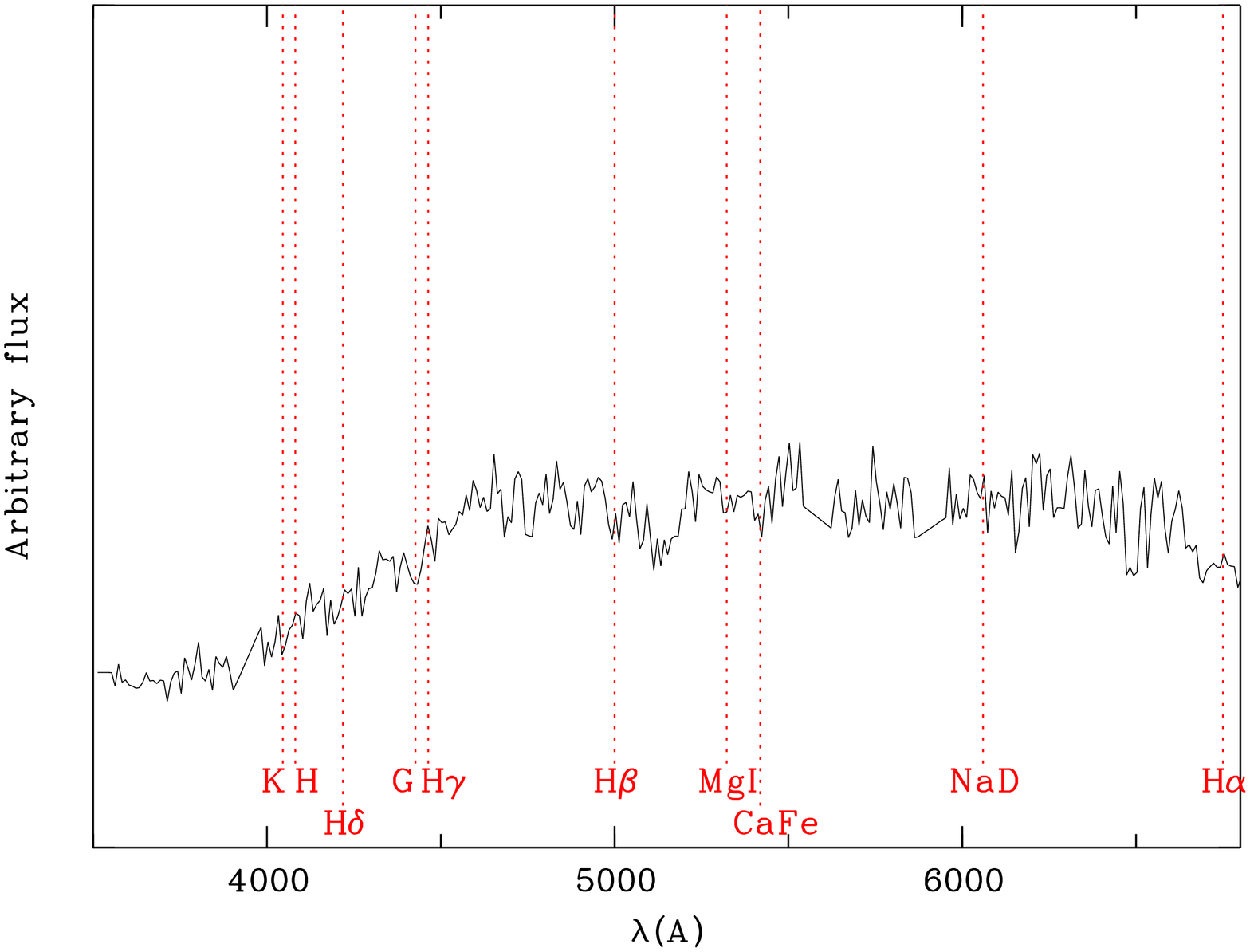,width=7cm,angle=270}}
\centering \mbox{\psfig{figure=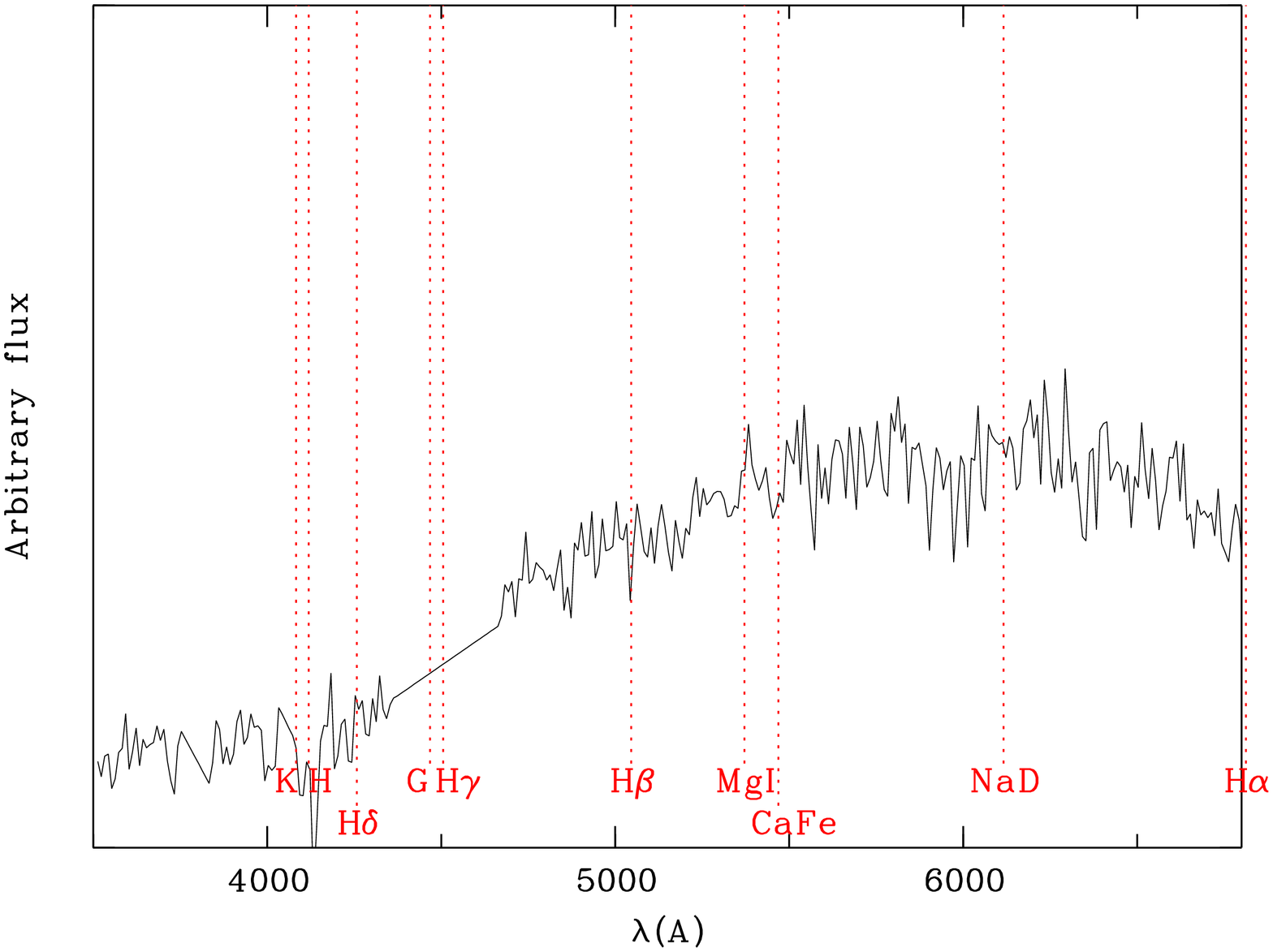,width=7cm,angle=270}}
\centering \mbox{\psfig{figure=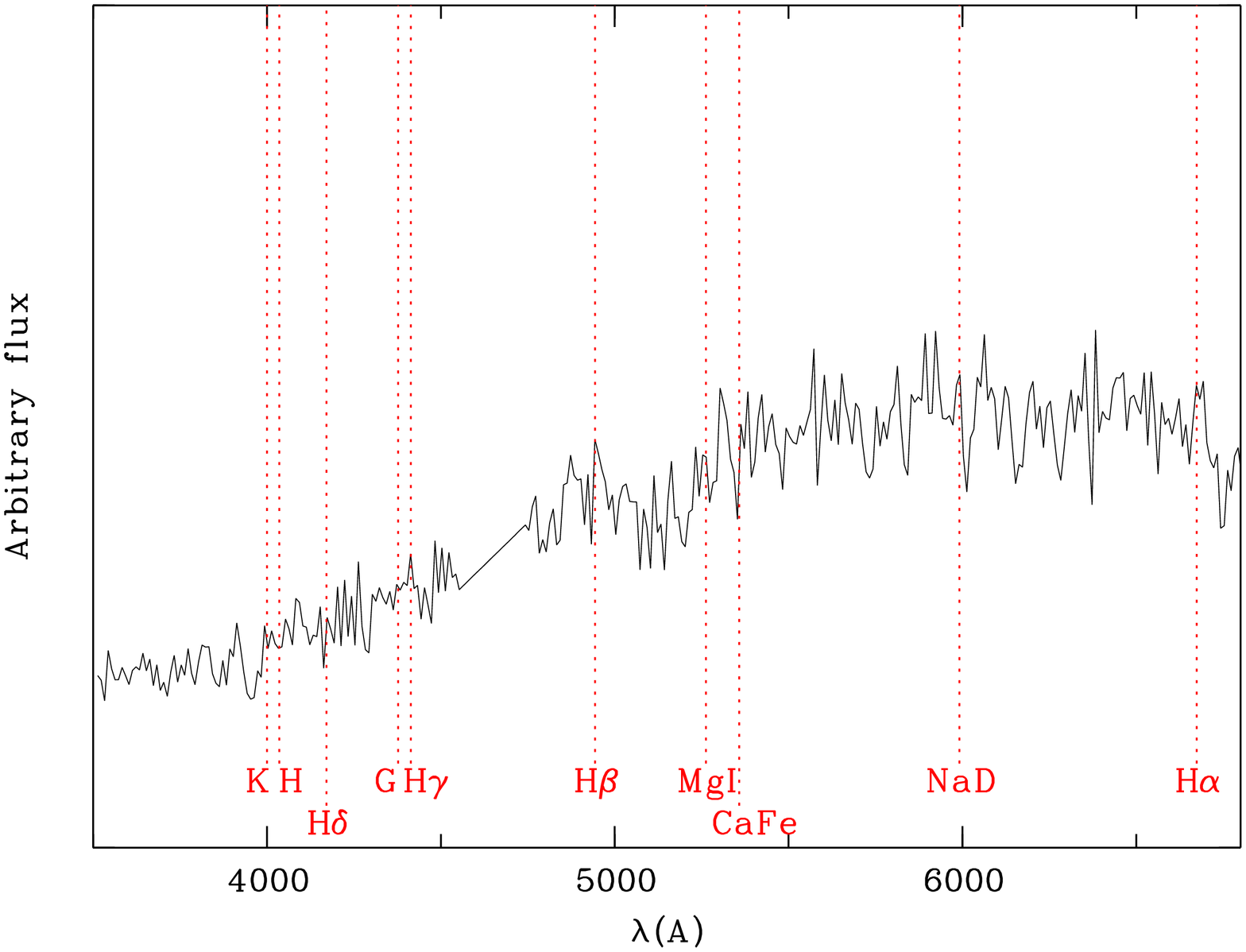,width=7cm,angle=270}}
\centering \mbox{\psfig{figure=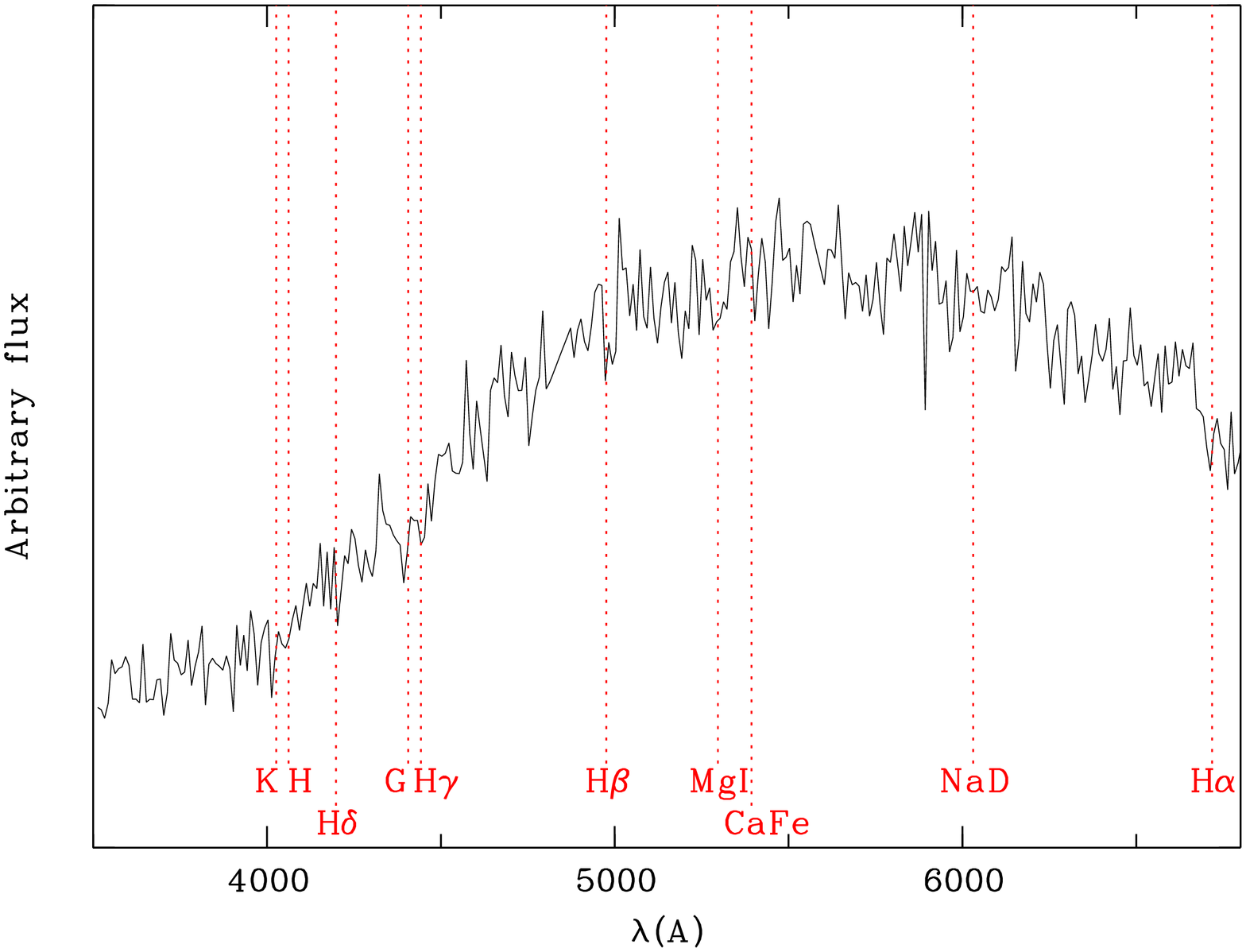,width=7cm,angle=270}}
\centering \mbox{\psfig{figure=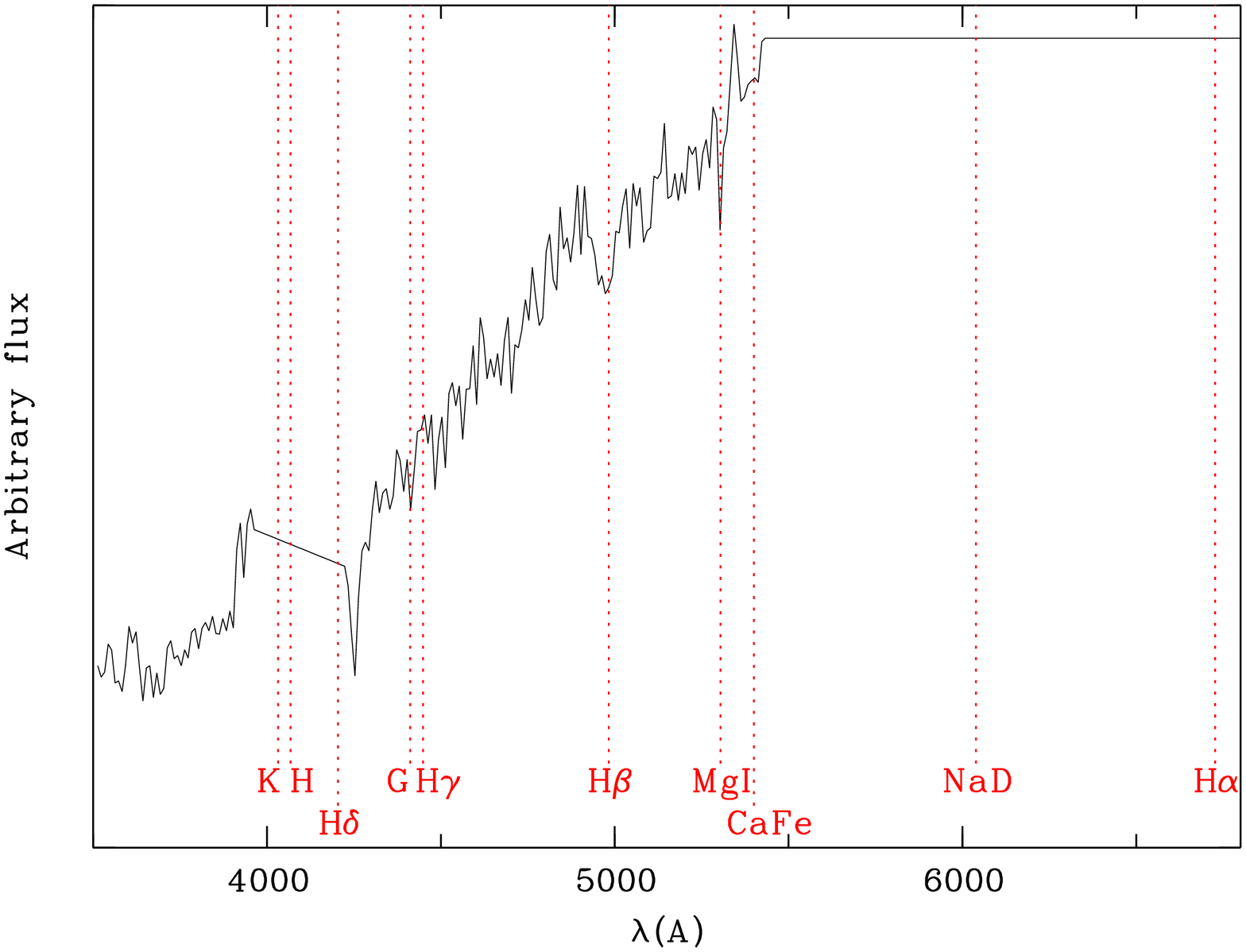,width=7cm,angle=270}}
\caption[]{From top to bottom, spectra of the 14783, 16601, 22156, 30341, and 36392 UCD candidates. }  
\label{fig:UCD3}
\end{figure}

We examined the spectra of these five objects in order to check if they
have the typical characteristics of UCD galaxies. The five spectra in
Fig.~\ref{fig:UCD3} proved to be very similar to the Price et al. (2009)
objects, with no detectable emission line.  However, our spectra have a
too low signal to noise ratio to efficiently adjust spectral models for
these five galaxies and we will not push forward such an analysis.

\subsection{UCD candidates}

As such galaxies are probably produced via strong interaction processes,
we checked the location of our 5 candidates. Two are very close to NGC~4874
and all but one are located along the  NGC~4874 - NGC~4839 infall
direction, where interaction processes are the strongest. Our sample is
however rather small and we need more data in order to be able to draw firm
conclusions. We give our candidate sample in Table~\ref{UCD}. 

\begin{table*}
\caption{UCD candidates: general identification, J2000 coordinates, B, R, I
  magnitudes, and redshift.}
\begin{center}
\begin{tabular}{ccccccc}
\hline
Id &$\alpha$   &  $\delta$ & B & R & I & redshift \\
 &  (deg)    &  (deg)    &  (mag)  &  (mag)   & (mag)  & \\
\hline
22156 ACO1656:194.789+27.8238 &  194.789      &    27.8238      & 24.42 & 22.42 & 22.15 & 0.0169  \\ 
16601 ACO1656:194.757+27.7621 &  194.757      &    27.7621      & 24.03 & 22.69 & 22.18 & 0.0380  \\ 
30341 ACO1656:194.890+27.9066 &  194.890      &    27.9066      & 23.24 & 21.76 & 21.31 & 0.0236  \\ 
36392 ACO1656:194.834+27.9631 &  194.834      &    27.9631      & 23.12 & 21.38 & 20.94 & 0.0250  \\ 
14783 ACO1656:195.148+27.7418 &  195.148      &    27.7418      & 23.70 & 21.91 & 21.45 & 0.0285  \\ 
\hline
\end{tabular}
\label{UCD}
\end{center}
\end{table*}

More generally, we note that UCD candidates, even if all confirmed, are a minor
component of our spectroscopic sample. They represent at most 5$\%$ of the
observed dwarf Coma cluster galaxies. General conclusions drawn in this
paper therefore apply to normal dwarf galaxies rather than to UCD galaxies.

\section{Magnitude and colour distributions of Coma cluster galaxies}

\subsection{Luminosity function}

In order to compute a luminosity function (LF), we have to count
the number of galaxies part of the Coma cluster per unit magnitude and
per unit area. The most straightforward way to compute such a LF
would be to have spectroscopic information for all galaxies along the
line of sight. This is certainly an unreachable astronomer's dream for
dwarf galaxies (see also Fig.~\ref{fig:comp}). Therefore for a given
magnitude range we have to estimate the percentage of cluster members
inside the complete photometric catalog using the spectroscopic
sample. This allows us to directly estimate the number of Coma cluster
galaxy members, and then the LF.

Such a calculation assumes that the photometric sample is complete. This
is not a concern however, because we have shown in Adami et al. (2006a,
2008) that all bands were basically 100$\%$ complete at least down
to R=23, and this is the faint limit of our spectroscopy.
 
A more serious concern is the representativeness of the spectroscopic
catalog. This catalog has to be free from significant selection biases,
which could potentially favour (or not) cluster galaxy members versus
non cluster members. Such selection effects are likely to occur at least
partially for VIMOS targets because part of them were selected on the
basis of a photometric redshift technique in order to increase the
number of cluster members. Literature spectra also come from several
samples and the selection effects are a priori unpredictable. We
therefore study in detail these possible selection biases. 

\begin{figure}
\centering \mbox{\psfig{figure=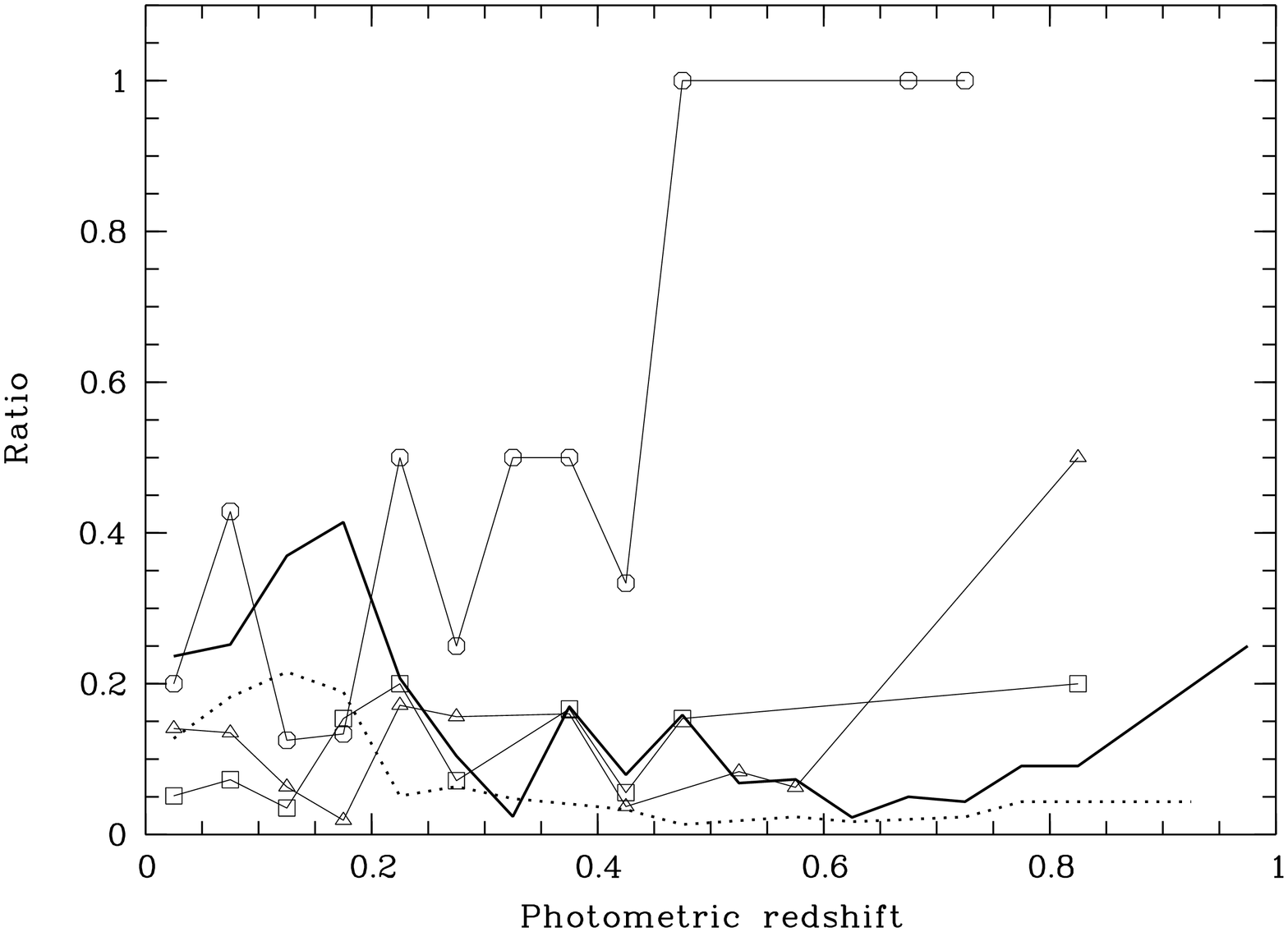,width=8cm,angle=270}}
\centering \mbox{\psfig{figure=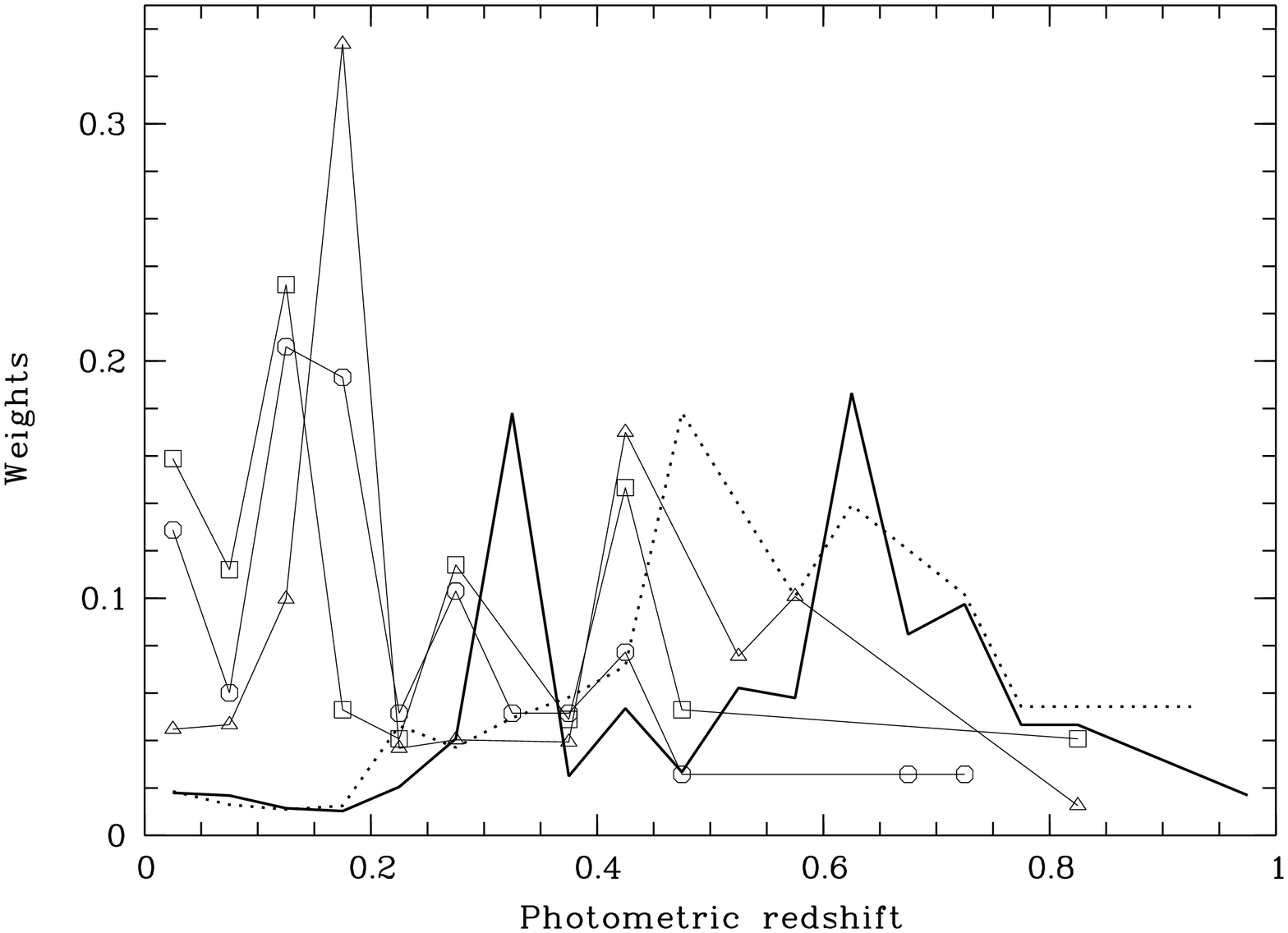,width=8cm,angle=270}}
\caption[]{Upper graph: direct ratio of the $photometric$ redshift
histogram of the $photometric$ sample to the $photometric$ redshift
histogram of the $spectroscopic$ sample. This is another way to show the
completeness of the spectroscopic sample compared to the photometric
redshift sample.

Lower graph: Correcting weights for the spectroscopic sample as a function
of $photometric$ redshift. Linked circles: R=[18;19], linked squares:
R=[19;20], linked triangles: R=[20;21], solid line: R=[21;22], dotted
line: R=[22;23].}  \label{fig:ponder}
\end{figure}

For each
magnitude bin, we compare the $photometric$ redshift histogram of the
$photometric$ sample with the $photometric$ redshift histogram of the
$spectroscopic$ sample (merging VIMOS and spectroscopic data from the
literature). Since the main bias comes from the photometric redshift
selection, this comparison allows us to quantify the selection
biases. If no bias is present, the ratio of these two histograms should
be constant as a function of $photometric$ redshift. The observed trends are
indeed relatively weak, showing however that we preferentially select 
low redshift objects for the
VIMOS sample and preferentially z$\geq$0.2 objects for the literature
sample. We therefore have to allocate a weight to each galaxy in order
to correct the Coma cluster member percentage. Considering the ratio of
the two previous histograms with $photometric$ redshift bins of 0.025,
we computed such a weight for each galaxy in our spectroscopic
sample. In a few words, galaxies preferentially selected in the
spectroscopic sample by selection effects are allocated a smaller
weight. Fig.~\ref{fig:ponder} shows these weights (and the direct value of the
ratio of the $photometric$ redshift histogram of the
$photometric$ sample to the $photometric$ redshift histogram of the
$spectroscopic$ sample) as a function of
$photometric$ redshift for the spectroscopic sample and for five
magnitude bins.

\begin{figure}
\centering \mbox{\psfig{figure=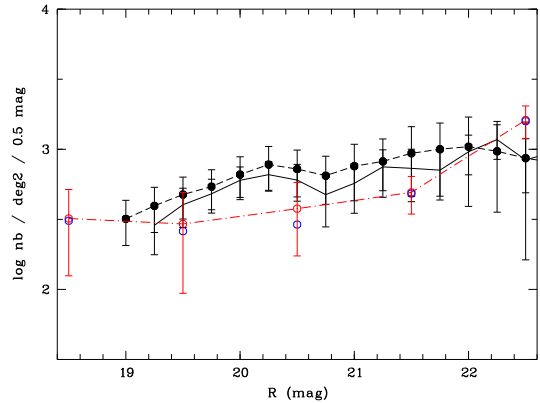,width=8cm,angle=270}}
\caption[]{Coma cluster LFs. Filled dashed-linked circles with error
bars: LF from pure photometric redshift technique (Adami et
al. 2008). Continuous and dotted lines with error bars: statistical subtraction
technique (Adami et al. 2007a). Open circles: present spectroscopic
technique (blue non linked: non corrected for selection biases, red dot-dashed
connected: corrected for selection biases).}  \label{fig:fdl}
\end{figure}

With these weights in hand, we are now able to compute corrected
percentages of Coma cluster members using our spectroscopic catalog, and
then LFs. We are still limited by the relatively low number of galaxies
with spectroscopic redshifts and in order to minimize the error bars we
only perform such a calculation in the whole area covered by the VIMOS
spectroscopy. The price to pay is that we cannot study environmental
effects on the LF with this technique. Fig.~\ref{fig:fdl} shows this LF,
as well as the previous estimations from Adami et al. (2007a, 2008)
based on statistical subtraction and on photometric redshift estimates
of the cluster membership in nearly the same area (namely subregions 4,
6, 7, 8, 11, 12, plus half of regions 9 and 16 of Adami et
al. 2008). Error bars from the spectroscopic estimates are
Poissonian. We clearly see the general agreement within the respective
error bars, even if these errors remain quite large. 
The R=[20,21] interval shows the largest differences, the
spectroscopic estimate being lower than the statistical subtraction and
pure photometric redshift estimates. This is easily understandable
because this bin is only poorly sampled by literature data and is not
yet filled by our VIMOS spectroscopy (which starts at R$\geq$21). This is
also likely to be the magnitude regime when the newly discovered
background groups contribute, implying a slightly lower LF for the Coma
cluster itself {\it in this specific region of the sky}. The R$\geq$21
range shows a better agreement with previous estimates. 

We also have a good agreement between the mean $\alpha$ slopes of the LF
for the three techniques (computed on the magnitude interval shown in
Fig.~\ref{fig:fdl}). We stress here that the slopes are not constant
(the points are not perfectly aligned) so we only compute mean
tendencies. However, given the error bar size, a constant slope cannot
be excluded. The statistical subtraction leads to $\alpha=-1.33\pm
0.19$, the pure photometric redshift technique to $\alpha=-1.31\pm
0.17$, and the present spectroscopic technique to $\alpha=-1.41\pm
0.45$.  We also see that the correcting weights do not induce large
differences for the spectroscopic LF estimate, showing that selection
effects, even if present, are not major. All these results put our
previous results (Adami et al. 2007a,b, 2008) on firmer grounds.

\subsection{Color magnitude relation}

\begin{figure}
\centering \mbox{\psfig{figure=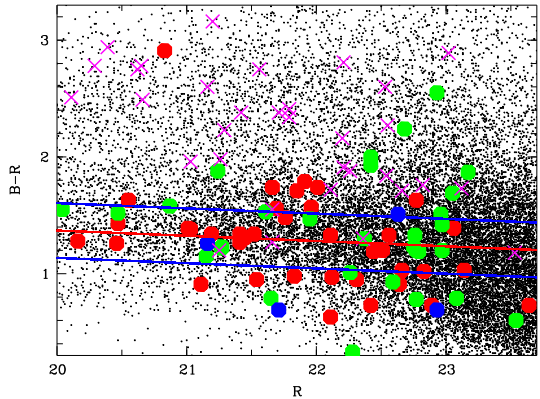,width=8cm,angle=270}}
\caption[]{B$-$R versus R band magnitude for galaxies (small black dots)
in our field of view. Considering the VIMOS sample only, pink crosses
are absorption line galaxies at z$\ge$0.2, red filled circles are Coma
cluster absorption line galaxies, green filled circles are Coma cluster
absorption+emission line galaxies, and blue filled circles are Coma
cluster emission line galaxies. The inclined red and blue lines are the
red sequence and its $\pm$2$\sigma$ uncertainty drawn from the Coma
cluster giant galaxies in Adami et al. (2006a).}  \label{fig:CMR_RS}
\end{figure}

We show in Fig.~\ref{fig:CMR_RS} the B$-$R versus R distribution of the
galaxies along the Coma cluster line of sight, comparing the photometric
sample and several VIMOS spectroscopic subsamples. Using the red
sequence (RS hereafter) defined by the giant galaxies from Adami et
al. (2006a), we see that the R$\geq$20 Coma members still statistically 
follow this relation. There is only a small percentage of outliers (e.g. 
the B$-$R$\sim$2.9 Coma galaxy). This percentage is easily explained by
 wrong redshifts (we recall that up to $\sim$10$\%$ of our VIMOS
redshifts could be wrong) or bad photometry.

Faint low surface brightness galaxies (Adami et al. 2006b) in Coma also
exhibit a similar RS compared to giant galaxies and we now generalize this
property to all Coma dwarf galaxies. Pure absorption line dwarf Coma
members exhibit a lower dispersion around the mean RS, but
emission+absorption line cluster galaxies (likely to be later type
objects) also seem to follow the RS.

In order to check that our photometry and our redshift determinations
are correct, we selected z$\ge$0.2 pure absorption line galaxies. These
likely early type objects (early type objects do not usually show
emission lines) are distant enough to show significant color evolution
compared to z=0, and even if they could form their own RS, they should
not overlap the RS of the Coma cluster giant galaxies. This is indeed verified
in Fig.~\ref{fig:CMR_RS}, the large majority of the z$\ge$0.2 pure absorption
line galaxies lying outside the Coma RS area.

\begin{figure}
\centering \mbox{\psfig{figure=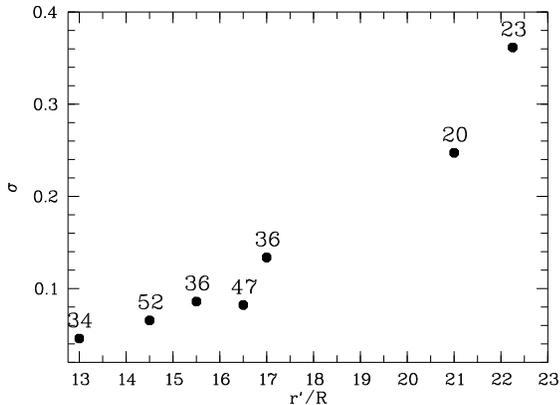,width=8cm,angle=270}}
\caption[]{Variation of the RS $\sigma$ (g'-r' versus r' for SDSS and
B-R versus R for VIMOS) as a function of magnitude (r' for SDSS and R
for VIMOS). SDSS data provide the points below a magnitude of 18, VIMOS
data above 18. The numbers of galaxies in each bin are also indicated
above each point.}  \label{fig:CMR_RS2}
\end{figure}

In order to compare our results for dwarf galaxies with trends observed
for giant galaxies, we extracted all SDSS DR7 z=[0.011;0.035] galaxies
in a circle of 25 arcmin centered on the Coma cluster centre. We
manually discarded all galaxies showing emission lines to have the same
selection as for the VIMOS z=[0.011;0.035] pure absorption line
galaxies. We show in Fig.~\ref{fig:CMR_RS2} the variation of the
dispersion $\sigma$ of the resulting RS as a function of magnitude
(g'-r' versus r' for SDSS and B-R/R for VIMOS). The faint dwarf pure
absorption line galaxies presently considered do not form a very compact
RS, as opposed to the Coma giant galaxies. We also find a continuous
increase of the RS $\sigma$ as a function of magnitude. This tendency
was already suspected by Adami et al. (2000) using shallower
spectroscopy and is probably related to the various origins of these
dwarf galaxies: primordial cluster galaxies or debris coming from a wide
range of progenitor galaxy types. Another possibility is downsizing, the
formation time of galaxies being regulated by their mass (Cowie et
al. 1996).

\section{A comprehensive picture of the Coma cluster}

We have presented new spectroscopic data covering part of the central
regions of the Coma cluster in the unprecedented magnitude range
$21\leq$R$\leq 23$. This range corresponds to the absolute magnitude
range $-14\leq {\rm M_R} \leq -12$, and therefore reaches dwarf galaxies, some
of them barely more massive than globular clusters. Combined with a
compilation of all redshifts available in the literature, this leads to
the most complete and deepest redshift catalogue presently available for
the Coma cluster, and to our knowledge for any cluster.

These data have allowed us to confirm the substructures previously
identified in Coma by Adami et al. (2005a), and to add three more
substructures.

We detect a large number of groups behind Coma, in particular a large
structure at z$\sim$0.5 and the SDSS Great Wall (Gott et al. 2005). We
also detect a large and very young previously unknown structure at
z$\sim$0.054, which may be located behind Coma more or less on the line
of sight, and that we named the background massive group (BMG). This structure
could be a major component of an unknown filament joining Coma to the SDSS
Great Wall. 

These structures allow to account for the mass maps derived from the weak
lensing analysis by Gavazzi et al. (2009).

By analyzing the spectral characteristics of our sample we show that
Coma cluster dwarf galaxies are old when red and young when blue. We
also find, by considering emission and absorption line characteristics,
that Coma cluster dwarf galaxies are old when showing absorption lines
and young when showing emission lines. The same trend is observed as a
function of stellar mass: red or absorption line galaxies have larger
stellar masses than blue or emission line galaxies.  However, the dust
content and stellar metallicities do not significantly
vary with spectral characteristics or colors. 

X-ray substructures do not seem to have any influence on the Coma cluster
galaxy properties. The magnitudes of the objects are not correlated with any
galaxy properties (besides the stellar mass).
However we put  in evidence spectral differences between R$\geq$22 and R$\leq$22
Coma cluster galaxies, the brightest showing less prominent absorption lines
(e.g. H$\&$K). This trend is less clear for field galaxies which are similar to
R$\geq$22 Coma galaxies. This suggests, along with the galaxy orbit features,
that part of the faint Coma galaxies could have been recently injected from 
the field following e.g. the NGC~4911 group infall. This could be put in perspective
  with the fact that red dwarf galaxies are generally very rare in the field
  (unless associated with bright objects) and that blue dwarf galaxies are
  generally very rare in clusters (because of gas stripping processes). In
  Coma, even if blue dwarf galaxies represent only 25$\%$ of the total sample,
their number is not negligible and is probably explained by the continuous
infall of field galaxies. 

Our spectroscopic data also allow us to confirm our previous results
concerning the galaxy luminosity functions based on a statistical
background subtraction and on photometric redshifts (Adami et al. 2007a,
b, and 2008) down to R=23. We find however that the LF in the BMG region
is somewhat overestimated when spectroscopy is not considered. Our
spectroscopy is not deep enough to confirm our luminosity functions
between R=23 and 25, but having a relatively good agreement in R=[21,23]
suggests that the agreement is probably also good at fainter magnitudes
and that luminosity functions are extremely steep at R$\geq$23 ($\alpha
\sim -2$ from Adami et al. 2007b).

Although our sample is still limited, we show that the orbits of dwarf
galaxies are probably anisotropic and radial. Dwarf galaxies have a
roughly constant density profile in the considered region, agreeing with
the hypothesis that they could originate from field galaxies radially
falling into the cluster. In this scenario, at least part of the dwarf
galaxies observed in Coma have fallen onto the cluster along the
numerous cosmological filaments surrounding Coma. In this hypothesis,
the steep faint end slope of the galaxy luminosity function would be
linked to the cluster environment.

In a B$-$R versus R colour--magnitude diagram, dwarf galaxies also form
a red sequence which is very similar to that of the giant galaxies,
which are 10 magnitudes brighter. We therefore confirm the red sequence
drawn with low surface brightness galaxies, for which of course
spectroscopic information is only very sparse (only 7 Coma low surface
brightness galaxies from the Adami et al. 2006b sample have a spectrum). Note
however that the dispersion on either side of the red sequence is quite
large for dwarf galaxies, implying that these galaxies have not all
followed the same formation processes. 

We have now acquired a good knowledge of the overall structure and
ongoing processes in the Coma cluster. We would now like to analyze spectroscopically
the GLF in the outer regions of Coma.  It was previously found from
imaging data that this slope steepened with increasing distance from the
center (Lobo et al. 1997, Beijersbergen et al. 2002, Adami et al. 2008), an important
result to understand how the Coma cluster formed. However, such a
steepening could be at least partly due to structures located behind
Coma, and now requires to be confirmed spectroscopically.  We also plan
to acquire low resolution spectroscopy similar to the one presented in
this paper for very faint galaxies in infalling structures such as the
NGC~4839 group, which we are presently analyzing in X-rays.

The following step will be to obtain spectroscopic data for low
surface brightness galaxies (including high resolution spectra to
compute the galaxy dynamical mass), but this will probably have to wait for the
next generation of extremely large telescopes due to prohibitive
exposure times even with 10m-class telescopes.

\begin{acknowledgements}
The authors thank the referee for useful and constructive comments.
We are grateful to the CFHT and Terapix (for the use of QFITS, SCAMP
and SWARP) teams, and to
the French CNRS/PNG for financial support. MPU
also acknowledges support from NASA Illinois space grant
NGT5-40073 and from Northwestern University.
We gratefully thank the ESO staff for succeeding in observing 
such a northern target from the Paranal site.
One of us (RT) thanks Gary Mamon for general guidance and for help on some
critical NED/SDSS cross-identifications. 
\end{acknowledgements}


\begin{thebibliography}{}

\bibitem[]{} Adami  C.,  Biviano A., Mazure A.,  1998, A$\&$A 331, 439

\bibitem[]{} Adami  C.,  Ulmer M.P., Durret F., et al.,  2000, A$\&$A 353, 930

\bibitem[]{} Adami  C.,  Biviano A., Durret F., Mazure A.,  2005a, A$\&$A 443,  17

\bibitem[]{} Adami  C.,  Slezak E., Durret F., et al., ,  2005b, A$\&$A 429,  39 

\bibitem[]{} Adami  C.,  Picat  J.P.,  Savine  C., et al.,  2006a, A$\&$A 451,  1159

\bibitem[]{} Adami  C.,  Scheidegger R., Ulmer M.P., et al.,  2006b, A$\&$A 459, 679

\bibitem[]{} Adami  C.,  Durret F., Mazure A., et al.,  2007a, A$\&$A 462, 411

\bibitem[]{} Adami  C.,  Picat  J.P.,  Durret F., et al.,  2007b, A$\&$A 472, 749 

\bibitem[]{} Adami  C.,  Ilbert O., Pell\'o R., et al.,  2008, A$\&$A 491, 681

\bibitem[]{} Adami  C.,  Gavazzi R., Cuillandre J.C., et al.,  2009a, A$\&$A 493, 399

\bibitem[]{} Adami  C.,  Pell\'o R., Ulmer M.P., et al.,  2009b, A$\&$A 495, 407

\bibitem[]{} Andreon S., Cuillandre J.C., 2002, ApJ 569, 144

\bibitem[]{} Beijersbergen M., Hoekstra H., van Dokkum P.G., van der
Hulst T. 2002, MNRAS 329, 385

\bibitem[]{} Bergeron J., Boiss\'e P., 1991,  A$\&$A 243, 344

\bibitem[]{} Biviano A., 1998, Untangling Coma Berenices: A New Vision of an
  Old Cluster, Proceedings of the meeting held in Marseilles (France), June
  17-20, 1997, Eds.: Mazure, A., Casoli F., Durret F. , Gerbal D., Word
  Scientific Publishing Co Pte Ltd, p. 1. 

\bibitem[]{} Biviano A., Durret F., Gerbal D., et al., 1996, A$\&$A 311, 95

\bibitem[]{} Biviano A., Katgert P., 2004, A$\&$A 424, 779

\bibitem[]{} Bournaud F., Duc P. A., Masset F. 2003, A\&A, 411, L469

\bibitem[]{} Chabrier G., 2003, PASP 115, 763

\bibitem[]{} Chilingarian A., Cayatte V., Bergond G. 2008,  MNRAS 390, 906

\bibitem[]{} Covone G., Adami C., Durret F., et al., 2006, A\&A 460, 381

\bibitem[]{}  Cowie L.L., Songaila A., Hu E.M., Cohen J.G., 1996, AJ 112, 839

\bibitem[]{} Croft R. A. C., Weinberg D. H., Bolte M., et al.,  2002, ApJ 581 20

\bibitem[]{} Gavazzi R., Adami  C., Durret F., et al.,  2009, A$\&$A
	  498, L33

\bibitem[]{} Geller M. J., Diaferio A., Kurtz M. J., 1999, ApJ 517, L23

\bibitem[]{} Gott J.R.III, Juri\'c M., Schlegel D., et al., 2005, ApJ 624, 463

\bibitem[]{} Iglesias-P\'aramo J., Boselli A., Gavazzi G., Cortese L.,
V\'\i lchez J.M. 2003, A\&A 397, 421

\bibitem[]{} Jenkins L. P., Hornschemeier A. E., Mobasher B., Alexander D. M.,
  Bauer F. E., 2007, ApJ 666, 846

\bibitem[]{} Kacprzak G.G., Churchill C.W., Steidel C.C., Murphy M.T., Evans
  J.L., 2007, ApJ 662, 909

\bibitem[]{} Khare P., Kulkarni V.P., P\'eroux C., et al., 2007, A\&A 464, 487

\bibitem[]{} King I.R., 1962, AJ 67, 471

\bibitem[]{} Kron R.G., 1980, ApJS 43, 305

\bibitem[]{} Lamareille F., Brinchmann J., Contini T., et al., 2009, A$\&$A
  495, 53

\bibitem[]{} Ledoux C., Valls-Gabaud D., Reboul H., et al., 1999, A$\&$AS
 138, 109

\bibitem[]{} Le F\`evre O., Vettolani G., Paltani S. et al., 2004, A$\&$A
  428, 1043

\bibitem[]{} Le F\`evre O., Vettolani G., Garilli B., et al., 2005, A$\&$A
  439, 845

\bibitem[]{} Lobo C., Biviano A., Durret F. et al., 1997, A\&A 317, 385

\bibitem[]{} Lokas E.L., Mamon G.A., 2003, MNRAS 343, 401

\bibitem[]{} Milne M.L., Pritchet C.J., Poole G.B., et al., 2007, AJ 133, 177

\bibitem[]{} Neumann  D.M., Lumb D.H., Pratt G.W., Briel U.G., 2003, A$\&$A 400, 811 

\bibitem[]{} Phillipps S., Drinkwater M.J., Gregg M.D., Jones J.B., et
	  al., 2001, ApJ 560, 201

\bibitem[]{} Price J., Phillipps S., Huxor A., et al., 2009, MNRAS astro-ph: 0906.1123 

\bibitem[]{} Salim S., Charlot S., Rich R.M., et al., 2005, ApJ 619, L39

\bibitem[]{} Sarazin C.L., 1986, Rev. of Mod. Physics 58, 1

\bibitem[]{} Scodeggio M., Franzetti P., Garilli B., et al., 2005, PASP
  117, 1284

\bibitem[]{} Serna A. \& Gerbal D. 1996, A\&A 309, 65

\bibitem[]{} Smith R. J., Marzke R. O., Hornschemeier A. E., et al., 2008,
  MNRAS 386, L96

\bibitem[]{} Strigari L.E., Bullock J.S., Kaplinghat M., et al., 2008, 
Nature 454, 1096

\bibitem[]{} Terlevich A.I., Calswell N., Bower R.G., 2001, MNRAS 326, 1547

\bibitem[]{} Trentham N. 1998, MNRAS 293, 71

\bibitem[]{} Walcher C.J., Lamareille F., Vergani D., et al., 2008, A$\&$A
  491, 713

\bibitem[]{} White S. D. M., Rees M. J., 1978, MNRAS 183, 341

\bibitem[]{} White S. D. M., Frenk C. S., 1991, ApJ 379, 52

\bibitem[]{} Wolf M., 1901, Astron. Nachr. 155, 127



\end{thebibliography}
\end{document}